\documentclass[preprint2]{aastex}

\slugcomment{Astronomical Journal submitted on 2015-07-28/ accepted on 2016-02-22}

\shorttitle{Star formation in 3CR galaxies}
\shortauthors{Westhues et al.}

\begin{document}

\title{
  Star formation\altaffilmark{\star} in 
  3CR radio galaxies and quasars at z $<$ 1
}

\author{
  Christian Westhues\altaffilmark{~1,2}, 
  Martin Haas\altaffilmark{~1}, 
  Peter Barthel\altaffilmark{~3}, 
  Belinda J. Wilkes\altaffilmark{~4}, 
  S.P. Willner\altaffilmark{~4}, 
  Joanna Kuraszkiewicz\altaffilmark{~4},
  Pece Podigachoski\altaffilmark{~3},
  Christian Leipski\altaffilmark{~5},
  Klaus Meisenheimer\altaffilmark{~5},
  Ralf Siebenmorgen\altaffilmark{~6},  and Rolf Chini\altaffilmark{~1,7}
\altaffiltext{$\star$}{
{\it Herschel} is an ESA space observatory with science instruments provided by European-led Principal Investigator consortia and with important participation from NASA.
}
\altaffiltext{1}{
  Astronomisches Institut, Ruhr-Universit\"at Bochum,
  Universit\"atsstr. 150, D-44801 Bochum,
  Germany
} 
\altaffiltext{2}{email: christian.westhues@astro.rub.de 
}
\altaffiltext{3}{
  Kapteyn Astronomical Institute, University of Groningen, NL-9747 AD
  Groningen
}
\altaffiltext{4}{
  Harvard-Smithsonian Center for Astrophysics, Garden St. 60, Cambridge, MA 02138,
  USA
}
\altaffiltext{5}{
  Max-Planck-Institut f\"ur Astronomie (MPIA), K\"onigstuhl 17, D-69117 Heidelberg,
  Germany
}
\altaffiltext{6}{
European Southern Observatory, Karl-Schwarzschild-Str. 2,
  D-85748 Garching b. M\"unchen, Germany
}
\altaffiltext{7}{
  Instituto de Astronom\'{i}a, Universidad Cat\'{o}lica del
  Norte, Avenida Angamos 0610, Casilla 1280, Antofagasta, Chile
}
}  

\begin{abstract}
  Using the {\it Herschel Space Observatory} we have observed a
  representative sample of 87 powerful 3CR sources at redshift $z <
  1$. The far-infrared (FIR, 70-500~$\micron$) photometry is combined
  with mid-infrared (MIR) photometry from the {\it Wide-Field Infrared
    Survey Explorer (WISE)} and catalogued data to analyse the
  complete spectral energy distributions (SEDs) of each object from
  optical to radio wavelength.  To disentangle the contributions of
  different components, the SEDs are fitted with a set of templates to
  derive the luminosities of host galaxy starlight, dust torus
  emission powered by active galactic nuclei (AGN) and cool dust
  heated by stars. The level of emission from relativistic jets is
  also estimated, in order to isolate the thermal host galaxy 
  contribution. The new data are in line with the
  orientation-based unification of high-excitation radio-loud AGN, in
  that the dust torus becomes optically thin longwards of $30~\micron$.
  The low excitation radio galaxies and the MIR weak sources represent
  MIR- and FIR-faint AGN population different from the high-excitation
  MIR-bright objects; it remains an open question whether they are at
  a later evolutionary state or an intrinsically different
  population. The derived luminosities for host starlight 
  and dust heated by
  star formation are converted to stellar masses and star formation
  rates (SFR). The host-normalized SFR of the bulk of the 3CR sources
  is low when compared to other galaxy populations at the same
  epoch. Estimates of the dust mass yield a 1--100 times lower
  dust/stellar mass ratio than for the Milky Way, indicating that
  these 3CR hosts have very low levels of interstellar matter 
  explaining the low level of star formation.
  Less than 10\% of the 3CR
  sources show levels of star formation above those of the main
  sequence of star forming galaxies.
\end{abstract}


\section{Introduction}

In the current paradigm of AGN evolution, galaxy collisions and mergers lead to the genesis of powerful radio sources (\citealt{Heckman86}). Based on FIR studies with the Infrared Astronomical Satellite ({\it IRAS}) in the 1980s, the Palomar-Green (PG) quasars appear to be preceded or accompanied by violent dust-enshrouded starburst activity (\citealt{Sanders88, Sanders89, Rowan-Robinson95}). Refined Infrared Space Observatory ({\it ISO}) photometry in the 1990s indicates a potential evolution from FIR-bright to FIR-faint AGN states (see \citealt{Haas03}).

Searching for the unbeamed counterparts of the quasar population in the medium-redshift ($0.5<z<1$) sample from the Revised Third Cambridge Catalogue of Radio Sources (3CR), \cite{Barthel89, Barthel94} proposed the orientation-based unification scheme of quasars and high-excitation radio galaxies (HERGs). Consensus is growing that this scheme is basically valid for sources with high radio power ($P_{178\,{\rm {MHz}}} > 10^{28}~WHz^{-1}$). 

The sample is subdivided by the classification criteria based on radio and optical properties. In compact steep spectrum (CSS) sources the radio emission is restricted to regions of less than 20kpc. Fanaroff-Riley Class I (FRI) sources show edge-dimmed radio lobes, while in FRII sources the lobes are more bright at the edge. The type-1 sources have optical bright continua and broad emission lines and are called Broad-Line Radio Galaxies (BLRG) at low luminosity. The high luminosity Flat-Spectrum-Quasars (FSQs) show flat radio spectra in $F_{\nu}$ in contrast to Steep-Spectrum-Quasars (SSQs) with the dividing spectral index $\alpha = 0.5$ measured at a few GHz. Type-2 sources show only narrow emission lines and have weak optical continua, High-Excitation RGs (HERGs) have [\ion{O}{3}]/[\ion{O}{2}] $>$ 1, Low-Excitation RGs (LERGs) have [\ion{O}{3}]/[\ion{O}{2}] $<$ 1 ($\lambda_{[O\,II]}=3727~\mathring{A}$, $\lambda_{[O\,III]}=5007~\mathring{A}$).

The 3CR radio sources can be sub-divided into many different classes (e.g. quasars and radio-galaxies), and it has early been questioned by demographic arguments that every edge-brightened double-lobe FR\,II radio galaxy is a misaligned hidden quasar. At low-redshift ($z<0.5$) where the radio power of the 3CR sample reaches down to $P_{178\,{\rm{MHz}}} \approx 10^{26}~WHz^{-1}$, narrow-line radio galaxies outnumber the quasars and broad-line radio galaxies (BLRGs), mainly due to the contribution of low-excitation radio galaxies (LERGs) (\citealt{Laing83, Singal93}).
 
Based on mid-infrared (MIR) observations, with VISIR, ISOCAM \citep{vanderWolk10,Sieb04}
and Spitzer \citep{Ogle06}, the LERGs and a few HERGs are MIR-weak,
indicating that they either do not possess high accretion power
comparable to the MIR-strong HERGs and quasars/BLRGs or that they are
more strongly extincted. FIR observations may be able to discriminate
between the two scenarios, but in view of the expected faintness such
observations have not been performed so far; only a few dozen bright
HERGs and quasars/BLRGs, have been detected in the FIR with
ISO\footnote{Due to the large ISO beam (FWHM = 90$''$ at 160~$\micron$),
  the ``high'' flux of some 3C-sources, e.g. 3C\,20 and 3C\, 47,
  contains contribution from nearby sources revealed by our Herschel
  maps} as compiled by \cite{Haas04}.

In this work a sample of 87 sources from the 3CR catalogue (\citealt{Edge59,Bennet62, Laing83, Spinrad85}) is studied. With the Herschel Space Observatory (\citealt{Pilbratt10}) 
we measured the FIR/submm SEDs of the 3CR sources 
in two complementary proposals, one at redshift $1<z<3$ (PI: Barthel, \citealt{Barthel12,Podigachoski15a,Podigachoski15b}) and one at medium ($0.5<z<1$) and low ($z<0.5$) redshift (PI: Haas).

We here present sensitive Herschel PACS/SPIRE 
70--500~$\micron$ photometry 
of the representative 3CR sample at low and medium-redshift.
The FIR properties of this 3CR sample 
were already measured with the 
previous IR satellites 
(IRAS: \citealt{Heckman92,Heckman94,Hes95,Hoekstra97};
ISO: \citealt{vanBemmel00, Fanti00, Polletta00,Meisenheimer01, Andreani02,Haas04}
and the Spitzer Space Telescope: \citealt{Haas05, Ogle06, Cleary07}). The new FIR observations with Herschel benefit from the higher spatial resolution and sensitivity of the instruments.

We here analyse the full optical to radio SEDs, 
also combined with {\it WISE} 3--22~$\micron$ photometry. 
The purpose is to explore dust emission in the FIR for the most powerful 
radio-loud AGN, to provide constraints 
on the star forming activity and 
to investigate the evolutionary status of their host galaxies.

We adopt a standard $\Lambda$CDM cosmology 
($H_{\circ} = 73$\,km\,s$^{-1}$\,Mpc$^{-1}$, $\Omega_{\Lambda}=0.73$, 
and $\Omega_{m}=0.27$, \citealt{Spergel07}).  

\section{Sample}

\subsection{Medium-redshift sample $0.5 < z < 1$}
\label{sec_data_med_sample}

The sample properties for 3CR-sources at medium-redshifts, which were observed by Herschel in the two open time programs from the {\it OT1\_mhaas\_2} and {\it OT1\_jstevens\_1} proposals, are given in Table \ref{3C_int_lst}. From the 48 sources at $0.5 < z < 1$ in the 3CR catalogue a representative subset of 39 was observed. The sources are selected to be brighter than 10~Jy at a frequency of 178\,MHz (\citealt{Laing83}). The sources that were not observed with Herschel don't bias the remaining subsample because their types are well represented. For the observed sources MIR photometry and/or spectroscopy from the Spitzer Space Telescope can be found in \citealt{Ogle06} and \citealt{Cleary07}.
Sources in the 3CR catalogue but not observed by Spitzer are 2 FSQs (3C\,345, 3C\,454.3), 1 SSQ (3C\,275.1), 4 HERGs (3C\,34, 3C\,217, 3C\,247, 3C\,277.2) and 1 LERG (3C\,41). Two HERGs, 3C\,175.1 and 3C\,220.3, have been removed from the analysis. The former has insufficient ancillary data in the literature, and the latter acts as gravitational lens for a submillimetre galaxy at $z=2.2$ (\citealt{Haas14}).

Thus a sample of 37 representative sources has been analysed, which consists of 7 FSQs (6 of them CSS), 7 SSQs (one BLRG), 22 HERGs (4 CSS) and 1 LERG.

\subsection{Low-redshift sample $z < 0.5$}
\label{sec_data_low_sample}

The 3CR sample properties at low-redshifts are shown in Table \ref{3C_low_lst}. It contains 48 sources at $z<0.5$, of which 40 sources are included in the 3CR catalog (\citealt{Laing83}). From the \citealt{Spinrad85} version of 3CR-sources, which extends to lower declinations, 4 additional sources belong to this sample. Mainly taken from the {\it OT1\_mhaas\_2} proposal, the whole Herschel Science Archive (HSA) was searched for 3C-sources and the complete Herschel-observed list was collected, which were observed also in the {\it OT1\_pogle01\_1}, {\it OT1\_rmushotz\_1}, {\it OT1\_lho\_1} and {\it OT1\_dfarrah\_1} proposals.

For all of them Spitzer MIR data are available. 23 sources were observed in the flux limited sample of \cite{Ogle06} with $S_{178\,{\rm {MHz}}} > 15 Jy$. Four remaining sources from the Ogle sample were planned but not observed with Herschel. The rest of 25 sources were selected from samples already seen with Spitzer by \cite{Haas05} (also observed by the ISO satellite), by \cite{Cleary07} and \cite{Hardcastle09} (X-ray selection). 

For the analysis the sample is subdivided into 4 FSQs (thereof
3 BLRGs), 10 SSQs (thereof 6 BLRGs), 19 HERGs  and 15 LERGs. All but 6 sources (3 HERGs and 3 LERGs) in this sample are morphologically classified as FR\,II sources (\citealt{Fanaroff74}). 

\section{Data}
\begin{table*}
\scriptsize
  \caption{3CR sources $0.5 < z < 1$ observed with Herschel\label{3C_int_lst}}
\begin{tabular}{ccccccccc}
  \hline
  &&&&&&&&\\
  Name & RA [J2000] & Dec [J2000] & Redshift & $D_{\rm L}$ [Mpc] & Type & Proposal-ID\tablenotemark{a} & PACS-OBSID & SPIRE-OBSID\\
  &&&&&&&&\\
  \hline
  &&&&&&&&\\
    3C006.1&		00 16 31.1&  +79 16 50&	 0.8404& 5193& HERG &  1 & 1342262061/62 & \\
    3C022.0&		00 50 56.3&  +51 12 03&  0.9360& 5935& BLRG &  2 & 1342237866/67 & \\
    3C049.0&		01 41 09.1&  +13 53 28&	 0.6207& 3568& HERG\tablenotemark{b} &  1 & 1342261865/66 & \\
    3C055.0&		01 57 10.5&  +28 51 38&	 0.7348& 4392& HERG &  1 & 1342261794/95 & 1342261703\\
    3C138.0&		05 21 09.9&  +16 38 22&	 0.7590& 4578& QSR\tablenotemark{b}  &  1 & 1342267270/71 & 1342268340\\
    3C147.0&		05 42 36.1&  +49 51 07&	 0.5450& 3048& QSR\tablenotemark{b}  &  1 & 1342268972/73 & \\
    3C172.0&		07 02 08.3&  +25 13 53\tablenotemark{c}&	 0.5191& 2876& HERG &  1 & 1342268994/95 & \\
    3C175.0&		07 13 02.4&  +11 46 15&	 0.7700& 4665& QSR  &  1 & 1342269004/05 & \\ 
    3C175.1&		07 14 04.7&  +14 36 22&	 0.9200& 5820& HERG &  2 & 1342242694/95 & 1342230780\\
    3C184.0&		07 39 24.2&  +70 23 11\tablenotemark{c}&	 0.9940& 6406& HERG &  2 & 1342243742/43 & 1342229126\\
    3C196.0&		08 13 36.0&  +48 13 03&	 0.8710& 5436& QSR  &  1 & 1342254180/81 & \\
    3C207.0&		08 40 47.6&  +13 12 24&	 0.6806& 4009& QSR  &  1 & 1342254575/76 & \\
    3C216.0&		09 09 33.5&  +42 53 46&	 0.6699& 3929& QSR\tablenotemark{b}  &  1 & 1342254561/62 & 1342255115\\
    3C220.1&		09 32 39.6&  +79 06 32&	 0.6100& 3498& HERG &  1 & 1342254194-96 & \\
    3C220.3&		09 39 23.8&  +83 15 26\tablenotemark{c}&	 0.6800& 3997& LENS &  1 & 1342221818/19 & 1342254521\\
    3C226.0&		09 44 16.5&  +09 46 17\tablenotemark{c}&	 0.8177& 5030& HERG &  1 & 1342255958/59 & 1342255165\\
    3C228.0&		09 50 10.8&  +14 20 01&	 0.5524& 3106& HERG &  1 & 1342255462-64 & \\
    3C254.0&		11 14 38.7&  +40 37 20&	 0.7366& 4418& QSR  &  1 & 1342255900/01 & \\
    3C263.0&		11 39 57.0&  +65 47 49&	 0.6460& 3755& QSR  &  1 & 1342255428-30 & \\
    3C263.1&		11 43 25.1&  +22 06 56&	 0.8240& 5078& HERG &  1 & 1342255685-87 & \\
    3C265.0&		11 45 29.0&  +31 33 47\tablenotemark{c}&	 0.8110& 4978& HERG &  1 & 1342255485/86 & \\
    3C268.1&		12 00 24.5&  +73 00 46\tablenotemark{c}&	 0.9700& 6214& HERG &  2 & 1342245706/07 & 1342229628\\
           &                      &           &              &      & &  & 1342247316/17 & \\
    3C280.0&		12 56 57.8&  +47 20 20\tablenotemark{c}&	 0.9960& 6426& HERG &  2 & 1342233434/35 & 1342232704\\
    3C286.0&		13 31 08.3&  +30 30 33&	 0.8499& 5275& QSR\tablenotemark{b}  &  1 & 1342259326/27 & 1342259451\\
    3C289.0&		13 45 26.4&  +49 46 33&	 0.9674& 6196& HERG &  2 & 1342233495/96 & 1342232711\\
    3C292.0&		13 50 41.9&  +64 29 36\tablenotemark{c}&	 0.7100& 4218& HERG &  1 & 1342257595-97 & \\
    3C309.1&		14 59 07.6&  +71 40 20&	 0.9050& 5698& QSR\tablenotemark{b}  &  1 & 1342259354-57 & \\
    3C330.0&		16 09 34.9&  +65 56 38\tablenotemark{c}&	 0.5500& 3082& HERG &  1 & 1342261369/70 & \\
    3C334.0&		16 20 21.8&  +17 36 24&	 0.5551& 3118& QSR  &  1 & 1342261319/20 & 1342263861\\
    3C336.0&		16 24 39.1&  +23 45 12&	 0.9265& 5868& QSR  &  1 & 1342261324-27 & \\
    3C337.0&		16 28 52.5&  +44 19 07\tablenotemark{c}&	 0.6350& 3674& HERG &  1 & 1342261350-53 & \\
    3C340.0&		16 29 36.6&  +23 20 13\tablenotemark{c}&	 0.7754& 4702& HERG &  1 & 1342261321-23 & \\
    3C343.0&		16 34 33.8&  +62 45 36&	 0.9880& 6356& HERG\parbox{0.2cm}{\tablenotemark{b,c}}  &  2	& 1342234218/19 & \\
    3C343.1&		16 38 28.2&  +62 34 44&	 0.7500& 4511& HERG\tablenotemark{b} &  1 & 1342261364-66 & \\
    3C352.0&		17 10 44.1&  +46 01 29&	 0.8067& 4937& HERG &  1 & 1342256219-21 & \\
    3C380.0&		18 29 31.8&  +48 44 46&	 0.6920& 4081& QSR\tablenotemark{b}  &  1 & 1342257947/48 & \\
    3C427.1&		21 04 06.8&  +76 33 11&	 0.5720& 3230& LERG &  1 & 1342261377/78 & \\
    3C441.0&		22 06 04.9&  +29 29 20&	 0.7080& 4193& HERG &  1 & 1342221833-36 & \\
    3C455.0&		22 55 03.9&  +13 13 34&	 0.5430& 3026& HERG\parbox{0.2cm}{\tablenotemark{b,c}}  &  1 & 1342258014-17 & \\
    &&&&&&&&\\
    \hline
\end{tabular}
\tablenotetext{a}{1=OT1\_mhaas\_2, 2=OT1\_jstevens\_1}
\tablenotetext{b}{CSS}
\tablenotetext{c}{coordinates/classifiaction revised}
\end{table*}

\begin{table*}
\scriptsize
  \caption{3CR sources $z < 0.5$ observed with Herschel\label{3C_low_lst}}
\begin{tabular}{ccccccccc}
  \hline
  &&&&&&&&\\
  Name & RA [J2000] & Dec [J2000] & Redshift & $D_{\rm L}$ [Mpc] & Type & Proposal-ID\tablenotemark{a} & PACS-OBSID & SPIRE-OBSID\\
  &&&&&&&&\\
  \hline
  &&&&&&&&\\
3C020.0&00 43 09.2&+52 03 36\tablenotemark{b}&0.1740& 804&HERG& 1 &               & 1342265338 \\
3C031.0&01 07 24.9&+32 24 45&0.0170& 66&LERG\tablenotemark{c} & 3 & 1342224218/19 & 1342236245 \\
3C033.0&01 08 52.9&+13 20 14\tablenotemark{b}&0.0597& 252&HERG& 1 & 1342261863/64 &            \\
3C033.1&01 09 44.3&+73 11 57\tablenotemark{b}&0.1810& 842&BLRG& 1 & 1342261944-46 &            \\
3C035.0&01 12 02.3&+49 28 36\tablenotemark{b}&0.0670& 286&HERG& 1 & 1342261413-16 &            \\
3C047.0&01 36 24.4&+20 57 28\tablenotemark{b}&0.4250&2252&QSR & 1 &               & 1342261707 \\
3C048.0&01 37 41.3&+33 09 35&0.3670&1892&QSR\tablenotemark{d} & 1 &               & 1342261702 \\
3C079.0&03 10 00.1&+17 05 59\tablenotemark{b}&0.2559&1244&HERG& 1 & 1342262229/30 &            \\
3C098.0&03 58 54.4&+10 26 03&0.0305& 126&HERG& 1 & 1342267198/99 &            \\
3C109.0&04 13 40.4&+11 12 15\tablenotemark{b}&0.3056&1529&BLRG& 1 & 1342267272/73 & 1342266668 \\
3C111.0&04 18 21.3&+38 01 36&0.0485& 205&BLRG& 4 & 1342239439/40 & 1342229105 \\
3C120.0&04 33 11.1&+05 21 16&0.0330& 138&BLRG& 4 & 1342241955/56 & 1342239936 \\
3C123.0&04 37 04.4&+29 40 14&0.2177&1037&LERG& 1 & 1342267256-59 &            \\
3C153.0&06 09 32.5&+48 04 15&0.2769&1367&LERG& 1 & 1342267224-27 &            \\
3C171.0&06 55 14.8&+54 08 57\tablenotemark{b}&0.2384&1152&HERG& 1 & 1342267228/29 &            \\
3C173.1&07 09 18.2&+74 49 32\tablenotemark{b}&0.2921&1453&LERG& 1 & 1342265540/41 &            \\
3C192.0&08 05 35.0&+24 09 50&0.0597& 260&HERG& 1 & 1342254172/73 &            \\
3C200.0&08 27 25.4&+29 18 45&0.4580&2475&LERG& 1 & 1342254174-77 &            \\
3C219.0&09 21 08.6&+45 38 57&0.1747& 815&BLRG& 1 & 1342254559/60 &            \\
3C234.0&10 01 49.5&+28 47 09&0.1849& 870&HERG& 1 & 1342255459    & 1342255182 \\
3C236.0&10 06 01.8&+34 54 10\tablenotemark{b}&0.1005& 449&LERG& 3 & 1342246697/98 & 1342246613 \\
       &          &         &      &    &    & 1 & 1342270912/13 &            \\
3C249.1&11 04 13.9&+76 58 58\tablenotemark{b}&0.3115&1566&QSR & 5 & 1342221763-66 & 1342229630 \\
3C268.3&12 06 24.7&+64 13 37&0.3717&1928&BLRG\tablenotemark{d} & 1 & 1342255424/25 &            \\
3C273.0&12 29 06.7&+02 03 09&0.1583& 734&QSR & 6 &               & 1342234882 \\
3C274.1&12 35 26.7&+21 20 35&0.4220&2246&HERG& 1 & 1342258032-35 &            \\
3C285.0&13 21 17.9&+42 35 15&0.0794& 349&HERG& 1 & 1342258514/15 & 1342256880 \\
3C300.0&14 22 59.8&+19 35 37\tablenotemark{b}&0.2700&1331&HERG& 1 & 1342262509-12 &            \\
3C305.0&14 49 21.6&+63 16 14&0.0416& 177&HERG\tablenotemark{c}& 3 & 1342223959/60 & 1342234915 \\
3C310.0&15 04 57.1&+26 00 58\tablenotemark{b}&0.0538& 233&LERG\tablenotemark{c}& 3 & 1342235116/17 & 1342234778 \\
3C315.0&15 13 40.1&+26 07 31&0.1083& 484&HERG\tablenotemark{c}& 3 & 1342224636/37 & 1342234777 \\
3C319.0&15 24 04.9&+54 28 06\tablenotemark{b}&0.1920& 903&LERG& 1 & 1342231879-82 &            \\
3C321.0&15 31 43.5&+24 04 19&0.0961& 426&HERG& 1 &               & 1342261679 \\
3C326.0&15 52 09.1&+20 05 24\tablenotemark{b}&0.0895& 395&LERG& 3 & 1342248732/33 & 1342238327 \\
3C341.0&16 28 04.0&+27 41 39\tablenotemark{b}&0.4480&2406&HERG& 1 & 1342261328/29 &            \\
3C349.0&16 59 28.9&+47 02 55\tablenotemark{b}&0.2050& 970&HERG& 1 & 1342261354/55 &            \\
3C351.0&17 04 41.4&+60 44 30\tablenotemark{b}&0.3719&1927&QSR & 5 & 1342232428-31 & 1342229147 \\
3C381.0&18 33 46.3&+47 27 03&0.1605& 737&BLRG& 1 & 1342261360/61 &            \\
3C382.0&18 35 03.4&+32 41 47&0.0579& 246&BLRG& 1 & 1342256206/07 &            \\
3C386.0&18 38 26.2&+17 11 50\tablenotemark{b}&0.0169& 68&LERG\tablenotemark{c} & 3 & 1342231672/73 & 1342239789 \\
3C388.0&18 44 02.4&+45 33 30&0.0917& 401&LERG& 1 & 1342261356-59 &            \\
3C390.3&18 42 08.9&+79 46 17\tablenotemark{b}&0.0561& 239&BLRG& 1 & 1342221871/72 &            \\
3C401.0&19 40 25.0&+60 41 36\tablenotemark{b}&0.2011& 947&LERG& 1 & 1342256194-97 &            \\
3C424.0&20 48 12.0&+07 01 17&0.1270& 567&LERG& 3 & 1342233349/50 & 1342244149 \\
3C433.0&21 23 44.5&+25 04 28\tablenotemark{b}&0.1016& 445&HERG\tablenotemark{c}& 1 & 1342219391/92 &            \\
       &          &           &      &    &    & 3 & 1342232731/32 & 1342234675 \\
3C436.0&21 44 11.7&+28 10 19&0.2145&1016&HERG& 3 & 1342235316/17 & 1342234676 \\
       &          &           &      &    &    & 1 & 1342257734-37 &            \\
3C438.0&21 55 52.3&+38 00 28\tablenotemark{b}&0.2900&1435&LERG& 1 & 1342259246-49 &            \\
3C452.0&22 45 48.8&+39 41 16&0.0811& 349&HERG& 1 & 1342259368/69 &            \\
3C459.0&23 16 35.2&+04 05 18&0.2201&1045&BLRG& 3 & 1342237979/80 & 1342234756 \\
  &&&&&&&&\\
  \hline
\end{tabular}
\tablenotetext{a}{1=OT1\_mhaas\_2, 2=OT1\_jstevens\_1, 3=OT1\_pogle01\_1, 4=OT1\_rmushotz\_1, 5=OT1\_lho\_1, 6=OT1\_dfarrah\_1}
\tablenotetext{b}{coordinates revised}
\tablenotetext{c}{FR\,I}
\tablenotetext{d}{CSS}
\end{table*}                          

For some objects of the medium-redshift sample a revision of the coordinates given in NED was necessary. We checked the positions given by \cite{Laing83} and inspected WISE images. Positions from high resolution radio maps (\citealt{Mullin06,Haas14}) or positions seen with Chandra at 2--8keV were taken whenever available. 
For the low-redshift sample the coordinates were revised to match Willot's positions\footnote{http://3crr.extragalactic.info/cgi/database}. Core positions from high resolution radio maps from VLA observation by \cite{Gilbert04} were taken whenever available.
The revised coordinates are indicated by footnotes in Tables~\ref{3C_int_lst} and~\ref{3C_low_lst}. Also the classifications for 3C\,343 and 3C\,455 from NED were altered from QSR to HERG based on classification given by \cite{Veron-Cetty10}.

\subsection{Herschel PACS and SPIRE}
\label{sec_data_herschel}

The data were downloaded
from the Herschel Science Archive (HSA) within the framework of the Herschel Interactive
Processing Environment (HIPE version 11.1.0, \citealt{Ott10}). For source extraction the tool SourceExtractor from \cite{Bertin96} was used and
additional routines were developed in the Interactive Data Language
(IDL) using the IDL Astronomy Library (\citealt{Landsman93}). 

\subsubsection{Observations}
\label{subsec_data_herschel_observation}
For the Photoconducter Array Camera and Spectrometer (PACS) (\citealt{Poglitsch10}) the {\it Scan-Map} observational mode was chosen to observe the sources photometrically at $70/100/160~\micron$ (blue/green/red). In a single-scan two filters (blue-red or green-red) were observed simultaneously. Often a cross-scan was done, with two consecutive single-scans with different scan directions. With the first scan in blue-red and the second in green-red combination the whole spectral range of PACS is covered, and the double scan in the red filter was combined afterwards to reach a higher sensitivity. For some sources deeper imaging was achieved by the combination of multiple cross-scans. 

The Spectral and Photometric Imaging Receiver (SPIRE) (\citealt{Griffin10}) observes in three bands at $250/350/500~\micron$ (short/mid/long) at once. The {\it Small-Scan-Map} observational mode was chosen.
For the medium-redshift sample the OBSIDs for the 98 PACS scans
and 12 SPIRE scan-maps are shown in Table \ref{3C_int_lst}. The
low-redshift sample was observed in 118 PACS scans and 23 SPIRE maps,
OBSIDs are given in Table \ref{3C_low_lst}. 

\subsubsection{PACS-Reduction}
\label{subsec_data_herschel_pacs_reduction}

The reduction of the PACS scan-maps was done in two steps, as bright sources have to be masked during the high-pass filtering (see \citealt{Popesso12}). In the first step a preliminary image is generated, which is then used to determine the positions for masking with SourceExtractor (only detections with 9 pixels above a 3$\sigma$ threshold are masked).

To minimize correlated noise and to get a good signal-to-noise ratio a pixel fraction of 0.6 and pixel sizes of 1.1$''$ , 1.4$''$ and 2.1$''$ for the 70~$\micron$, 100~$\micron$ and 160~$\micron$ band was chosen. Additionally the high-pass filter radius was set to 10, 15 and 20 readouts. The multiple scans were then combined with the {\it mosaic} task in HIPE.

The aperture corrected flux was determined for the pointlike sources in the frame. The target source was assumed to be that closest to and within 7$''$ of the known source position (as listed in Tables \ref{3C_int_lst} and \ref{3C_low_lst}). Images of size $2\arcmin \times 2\arcmin$ are shown in Appendix \ref{app_maps}.

We derived the photometric uncertainty as follows: Every frame comes with a coverage map which was used to generate 500 random positions on the map, where the coverage is greater than 75\% of its maximum. At these positions the HIPE routine $annularSkyAperturePhotometry$ was used to perform aperture photometry with the background calculated in an annulus. Values for aperture and annulus radii (recommended for fluxes $< 500$mJy) given in Table \ref{radii} follow the Herschel Webinar for ``PACS Point Source Photometry'' by Paladini\footnote{\url{https://nhscsci.ipac.caltech.edu/pacs/docs/Webinars/2012-07-13/Paladini.pdf}}.
The Gaussian dispersion of the 500 aperture-corrected fluxes was adopted as the $1\sigma$ uncertainty for each map (see also \citealt{Leipski13}), and is listed in Tables \ref{3C_flux_int} and \ref{3C_flux_low}. Where no sources could be detected a 3$\sigma$ upper limit is given.

\begin{table}
\caption{PACS Aperture and Annulus radii}
\label{radii}
\centering
\begin{tabular}{lccc}
\hline&&&\\
 & 70~$\micron$ & 100~$\micron$ & 160~$\micron$ \\
&&&\\
\hline&&&\\
Aperture radius [''] & 5.5 & 5.6 & 10.5 \\
Annulus inner radius [''] & 20 & 20 & 24 \\
Annulus outer radius [''] & 25 & 25 & 28 \\
Aperture correction & 0.61 & 0.57 & 0.63\\
&&&\\
\hline
\end{tabular}
\end{table}

\begin{table*}  
  \scriptsize
  \caption{3CR sources $0.5 < z < 1$ PACS and SPIRE flux densities. 1$\sigma$ uncertainties are given in brackets, upperlimits are 3$\sigma$}
  \label{3C_flux_int}
  \centering
  \begin{tabular}{cccccccc}
  \hline
  &&&&&&&\\
  Name & Figure & $F_{70}$ [mJy] & $F_{100}$ [mJy] & $F_{160}$ [mJy] & $F_{250}$ [mJy] & $F_{350}$ [mJy] &  $F_{500}$ [mJy]\\
  &&&&&&&\\
  \hline
  &&&&&&&\\
3C006.1 & Fig.~\ref{fig_seds_herg_mir_medium}        & \textless  14 & \textless  15 & \textless  30 &   &   &   \\
3C022.0 & Fig.~\ref{fig_seds_ssq}                    &  28( 3) &   & \textless  36 &   &   &   \\
3C049.0 & Fig.~\ref{fig_seds_herg_mir_strong}        &  16( 4) &  25( 5) &  27( 7) &   &   &   \\
3C055.0 & Fig.~\ref{fig_seds_herg_mir_strong}        &  90( 4) &  126( 4) &  123( 8) &  85( 9) &  42( 9) & \textless  42 \\
3C138.0 & Fig.~\ref{fig_seds_fsq1}                    &  47( 4) &  49( 5) &  58( 10) &  64( 11) &  70( 15) &  103( 14) \\
3C147.0 & Fig.~\ref{fig_seds_fsq1}                    &  59( 5) &  71( 5) &  72( 10) &   &   &   \\
3C172.0 & Fig.~\ref{fig_seds_herg_mir_weak}          & \textless  9 & \textless  14 & \textless  28 &   &   &   \\
3C175.0 & Fig.~\ref{fig_seds_ssq}                    &  26( 3) &  24( 4) & \textless  45 &   &   &   \\
3C175.1 & Fig.~\ref{fig_seds_nlrg_rejected}          & \textless  8 &   & \textless  22 & \textless  62 & \textless  50 & \textless  38 \\
3C184.0 & Fig.~\ref{fig_seds_herg_mir_strong}        &  12( 2) &   & \textless  24 & \textless  31 & \textless  22 & \textless  23 \\
3C196.0 & Fig.~\ref{fig_seds_ssq}                    &  24( 3) &  18( 4) & \textless  34 &   &   &   \\
3C207.0 & Fig.~\ref{fig_seds_fsq1}                    &  16( 4) &  21( 5) &  32( 8) &   &   &   \\
3C216.0 & Fig.~\ref{fig_seds_fsq1}                    &  79( 3) &  102( 5) &  143( 9) &  204( 8) &  229( 14) &  266( 22) \\
3C220.1 & Fig.~\ref{fig_seds_herg_mir_weak}          & \textless  8 & \textless  12 & \textless  18 &   &   &   \\
3C220.3 & Fig.~\ref{fig_seds_nlrg_rejected}          &  26( 3) &  99( 4) &  259( 11) &  452( 9) &  412( 8) &  259( 7) \\
3C226.0 & Fig.~\ref{fig_seds_herg_mir_strong}        &  44( 3) &  36( 5) & \textless  19 & \textless  31 & \textless  42 & \textless  42 \\
3C228.0 & Fig.~\ref{fig_seds_herg_mir_weak}          & \textless  8 & \textless  15 & \textless  23 &   &   &   \\
3C254.0 & Fig.~\ref{fig_seds_ssq}                    &  14( 4) & \textless  15 & \textless  27 &   &   &   \\
3C263.0 & Fig.~\ref{fig_seds_ssq}                    &  56( 4) &  43( 4) &  27( 8) &   &   &   \\
3C263.1 & Fig.~\ref{fig_seds_herg_mir_medium}        & \textless  8 & \textless  12 & \textless  21 &   &   &   \\
3C265.0 & Fig.~\ref{fig_seds_herg_mir_strong}        &  40( 4) &  47( 6) & \textless  41 &   &   &   \\
3C268.1 & Fig.~\ref{fig_seds_herg_mir_weak}          & \textless  8 &   & \textless  20 & \textless  38 & \textless  45 & \textless  23 \\
3C280.0 & Fig.~\ref{fig_seds_herg_mir_strong}        &  24( 3) &   & \textless  31 & \textless  38 & \textless  32 & \textless  40 \\
3C286.0 & Fig.~\ref{fig_seds_fsq2}                    &  36( 4) &  41( 4) &  59( 7) &  76( 11) &  84( 13) &  112( 15) \\
3C289.0 & Fig.~\ref{fig_seds_herg_mir_medium}        &  10( 3) &   & \textless  21 & \textless  32 & \textless  33 & \textless  35 \\
3C292.0 & Fig.~\ref{fig_seds_herg_mir_weak}          & \textless  7 & \textless  11 & \textless  18 &   &   &   \\
3C309.1 & Fig.~\ref{fig_seds_fsq2}                    &  43( 3) &  35( 4) &  39( 9) &   &   &   \\
3C330.0 & Fig.~\ref{fig_seds_herg_mir_medium}        & \textless  11 & \textless  20 & \textless  27 &   &   &   \\
3C334.0 & Fig.~\ref{fig_seds_ssq}                    &  69( 3) &  71( 4) &  51( 8) & \textless  40 & \textless  33 & \textless  67 \\
3C336.0 & Fig.~\ref{fig_seds_ssq}                    & \textless  8 & \textless  9 & \textless  23 &   &   &   \\
3C337.0 & Fig.~\ref{fig_seds_herg_mir_weak}          & \textless  6 & \textless  9 & \textless  13 &   &   &   \\
3C340.0 & Fig.~\ref{fig_seds_herg_mir_medium}        & \textless  7 & \textless  13 & \textless  24 &   &   &   \\
3C343.0 & Fig.~\ref{fig_seds_herg_mir_strong}        &  58( 2) &   &  73( 11) &   &   &   \\
3C343.1 & Fig.~\ref{fig_seds_herg_mir_medium}        &  11( 3) & \textless  17 & \textless  21 &   &   &   \\
3C352.0 & Fig.~\ref{fig_seds_herg_mir_weak}          & \textless  7 & \textless  12 &  19( 4) &   &   &   \\
3C380.0 & Fig.~\ref{fig_seds_fsq2}                    &  69( 4) &  94( 5) &  149( 9) &   &   &   \\
3C427.1 & Fig.~\ref{fig_seds_lerg}                   &   & \textless  6 & \textless  18 &   &   &   \\
3C441.0 & Fig.~\ref{fig_seds_herg_mir_medium}        &  8( 3) & \textless  11 & \textless  26 &   &   &   \\
3C455.0 & Fig.~\ref{fig_seds_herg_mir_medium}        & \textless  6 & \textless  8 & \textless  18 &   &   &   \\
  &&&&&&&\\
  \hline
  \end{tabular}

\end{table*}

\begin{table*}  
  \centering
  \scriptsize
  \caption{3CR sources $z < 0.5$ PACS and SPIRE flux densities. 1$\sigma$-errors are given in brackets, upperlimits are 3$\sigma$}
  \label{3C_flux_low}
  \begin{tabular}{cccccccc}
  \hline
  &&&&&&&\\
  Name & Figure & $F_{70}$ [mJy] & $F_{100}$ [mJy] & $F_{160}$ [mJy] & $F_{250}$ [mJy] & $F_{350}$ [mJy] &  $F_{500}$ [mJy]\\
  &&&&&&&\\
  \hline
  &&&&&&&\\
3C020.0 & Fig.~\ref{fig_seds_low_nlrg1}              &   &   &   &  58( 19) &  50( 16) & \textless  50 \\
3C031.0 & Fig.~\ref{fig_seds_low_lerg2}              &   &  867( 3) &  1347( 11) &  955( 16) &  408( 20) &  169( 20) \\
3C033.0 & Fig.~\ref{fig_seds_low_nlrg1}              &  161( 4) &  194( 4) &  179( 8) &   &   &   \\
3C033.1 & Fig.~\ref{fig_seds_low_blrg}               &  38( 4) &  31( 3) & \textless  39 &   &   &   \\
3C035.0 & Fig.~\ref{fig_seds_low_nlrg3}              & \textless  6 &  16( 3) & \textless  23 &   &   &   \\
3C047.0 & Fig.~\ref{fig_seds_low_ssq}                &   &   &   & \textless  41 & \textless  40 & \textless  23 \\
3C048.0 & Fig.~\ref{fig_seds_low_ssq}                &   &   &   &  311( 10) &  137( 13) &  62( 10) \\
3C079.0 & Fig.~\ref{fig_seds_low_nlrg1}              &  65( 4) &  49( 5) &  37( 10) &   &   &   \\
3C098.0 & Fig.~\ref{fig_seds_low_nlrg2}              &  41( 5) &  49( 4) & \textless  37 &   &   &   \\
3C109.0 & Fig.~\ref{fig_seds_low_blrg}               &  158( 4) &  106( 4) &  62( 9) & \textless  37 & \textless  46 & \textless  46 \\
3C111.0 & Fig.~\ref{fig_seds_low_fsq}                &  242( 6) &   &  461( 24) &  577( 27) &  741( 26) &  876( 31) \\
3C120.0 & Fig.~\ref{fig_seds_low_fsq}                &  783( 8) &   &  1145( 20) &  634( 9) &  465( 13) &  439( 16) \\
3C123.0 & Fig.~\ref{fig_seds_low_lerg}               &  22( 2) &  10( 3) & \textless  32 &   &   &   \\
3C153.0 & Fig.~\ref{fig_seds_low_lerg}               & \textless  8 & \textless  10 & \textless  20 &   &   &   \\
3C171.0 & Fig.~\ref{fig_seds_low_nlrg1}              &  13( 4) & \textless  15 & \textless  34 &   &   &   \\
3C173.1 & Fig.~\ref{fig_seds_low_lerg}               & \textless  10 &  13( 4) & \textless  30 &   &   &   \\
3C192.0 & Fig.~\ref{fig_seds_low_nlrg3}              &  28( 4) &  33( 4) & \textless  24 &   &   &   \\
3C200.0 & Fig.~\ref{fig_seds_low_lerg}               & \textless  7 &  12( 3) & \textless  18 &   &   &   \\
3C219.0 & Fig.~\ref{fig_seds_low_blrg}               & \textless  11 & \textless  12 & \textless  29 &   &   &   \\
3C234.0 & Fig.~\ref{fig_seds_low_nlrg1}              &   &  87( 4) &  46( 13) & \textless  40 & \textless  47 & \textless  38 \\
3C236.0 & Fig.~\ref{fig_seds_low_lerg2}              &  55( 5) &  90( 2) &  120( 9) &  92( 13) &  81( 15) &  78( 14) \\
3C249.1 & Fig.~\ref{fig_seds_low_ssq}                &  64( 3) &  62( 3) &  45( 6) & \textless  39 & \textless  46 & \textless  31 \\
3C268.3 & Fig.~\ref{fig_seds_low_blrg}               &  22( 3) &  32( 4) & \textless  26 &   &   &   \\
3C273.0 & Fig.~\ref{fig_seds_low_fsq}                &   &   &   &  475( 16) &  683( 11) &  1062( 20) \\
3C274.1 & Fig.~\ref{fig_seds_low_nlrg3}              & \textless  6 & \textless  10 & \textless  19 &   &   &   \\
3C285.0 & Fig.~\ref{fig_seds_low_nlrg2}              &  222( 4) &  292( 5) &  307( 11) &  180( 20) &  74( 13) & \textless  44 \\
3C300.0 & Fig.~\ref{fig_seds_low_nlrg3}              & \textless  6 & \textless  8 & \textless  19 &   &   &   \\
3C305.0 & Fig.~\ref{fig_seds_low_nlrg2}              &   &  381( 4) &  502( 9) &  254( 22) &  118( 17) & \textless  64 \\
3C310.0 & Fig.~\ref{fig_seds_low_lerg2}              &   &  23( 1) &  38( 3) &  \textless 30 & \textless  43 & \textless  41 \\
3C315.0 & Fig.~\ref{fig_seds_low_nlrg3}              &   &  31( 3) &  36( 9) & \textless  24 & \textless  27 & \textless  27 \\
3C319.0 & Fig.~\ref{fig_seds_low_lerg}               & \textless  7 & \textless  11 &  31( 8) &   &   &   \\
3C321.0 & Fig.~\ref{fig_seds_low_nlrg1}              &   &   &   &  287( 16) &  108( 13) & \textless  40 \\
3C326.0 & Fig.~\ref{fig_seds_low_lerg2}              &  6( 2) &  15( 1) &  19( 4) & \textless  30 & \textless  20 & \textless  23 \\
3C341.0 & Fig.~\ref{fig_seds_low_nlrg1}              &  28( 3) & \textless  11 & \textless  23 &   &   &   \\
3C349.0 & Fig.~\ref{fig_seds_low_nlrg2}              & \textless  11 & \textless  12 & \textless  23 &   &   &   \\
3C351.0 & Fig.~\ref{fig_seds_low_ssq}                &  172( 3) &  156( 4) &  89( 10) & \textless  60 & \textless  81 & \textless  34 \\
3C381.0 & Fig.~\ref{fig_seds_low_blrg}               &  35( 4) &  34( 5) &  39( 10) &   &   &   \\
3C382.0 & Fig.~\ref{fig_seds_low_fsq}                &  76( 4) &  96( 4) &  97( 11) &   &   &   \\
3C386.0 & Fig.~\ref{fig_seds_low_lerg2}              &   &  66( 3) &  81( 9) & \textless  90 & \textless  43 & \textless  43 \\
3C388.0 & Fig.~\ref{fig_seds_low_lerg2}              & \textless  7 & \textless  9 & \textless  16 &   &   &   \\
3C390.3 & Fig.~\ref{fig_seds_low_blrg}               &  157( 4) &  110( 4) &  51( 8) &   &   &   \\
3C401.0 & Fig.~\ref{fig_seds_low_lerg}               & \textless  8 & \textless  9 & \textless  18 &   &   &   \\
3C424.0 & Fig.~\ref{fig_seds_low_lerg}               &   &  7( 1) & \textless  13 & \textless  22 & \textless  39 & \textless  28 \\
3C433.0 & Fig.~\ref{fig_seds_low_nlrg1}              &  294( 4) &  288( 2) &  226( 6) &  120( 22) & \textless  51 & \textless  46 \\
3C436.0 & Fig.~\ref{fig_seds_low_nlrg3}              &  16( 3) &  30( 2) &  37( 5) & \textless  51 & \textless  30 & \textless  40 \\
3C438.0 & Fig.~\ref{fig_seds_low_lerg}               & \textless  7 & \textless  11 & \textless  42 &   &   &   \\
3C452.0 & Fig.~\ref{fig_seds_low_nlrg2}              &  38( 4) &  36( 5) &  27( 8) &   &   &   \\
3C459.0 & Fig.~\ref{fig_seds_low_blrg}               &   &  584( 4) &  549( 11) &  284( 15) &  115( 11) &  54( 16) \\
  &&&&&&&\\
  \hline
  \end{tabular}

\end{table*}

\subsubsection{SPIRE-Reduction}
\label{subsec_data_herschel_spire_reduction}

For the photometry of the SPIRE observations we followed the steps of the ``Recipe for SPIRE Photometry''\footnote{http://herschel.esac.esa.int/hcss-doc-13.0/load/spire\_drg/html/ch06s07.html}. The recommended algorithm for point source photometry is the time-line-fitter, {\it sourceExtractorTimeline} in HIPE. To determine the positions of sources in the SPIRE maps, we used the level2 products of the observations in the HSA. A source list was generated within HIPE by {\it sourceExtractorSussextractor}. The coordinates of these sources were then used to perform the fitting in the timeline data on the level1 products in the HSA. The nearest source within 30$''$ to the coordinates given in Table \ref{3C_int_lst} and \ref{3C_low_lst} was identified as the 3CR target.

As for the uncertainty determination for the PACS observations, we used 500 randomly generated positions on the SPIRE maps, centered within 23 pixels of the center ($138''$/$230''$/$322''$ for $250~\micron$/$350~\micron$/$500~\micron$). At these positions the photometry was carried out in the same manner as for the sources. The dispersion of the distribution again gave the 1$\sigma$ uncertainty or 3$\sigma$ upper limits, which are given in Tables \ref{3C_flux_int} and \ref{3C_flux_low}. Images of size $4\arcmin \times 4\arcmin$ are shown in Appendix \ref{app_maps}.

\subsection{ISOCAM and Spitzer}
\label{sec_data_spitzer}

We combined the SEDs in the MIR with data from the {\it Infrared Space
  Observatory} (\citealt{Kessler96}) and {\it Spitzer Space Telescope}
(\citealt{Werner04}). We used photometric imaging observations with
ISOCAM \citep{Cesarsky96} by \cite{Sieb04} and spectra taken with the
Infrared Spectrograph (IRS, \citealt{Houck04}) from $5.2$ to $38~\micron$, which were extracted by the CASSIS (\citealt{Lebouteiller11})
and newly stitched and scaled by the IDEOS project
(\citealt{Spoon12}). From previous analysis of the spectra
(\citealt{Ogle06}) flux densities at $7$ and $15~\micron$ restframe are
included. Also photometric Spitzer data from the Multiband Imaging
Photometer (MIPS, \citealt{Rieke04}) at $24~\micron$ are used
(\citealt{Shi05,Cleary07,Hardcastle09,Fu09,Dicken10,Shang11}).

\subsection{2MASS and WISE}
\label{sec_data_2mass_wise}

We queried the {{\it wise\_allwise\_p3as\_psd} data release from the Wide-field Infrared Survey Explorer (WISE, \citealt{Wright10}) with the IDL {\it query\_irsa\_cat} routine around $4"$ of the estimated positions for the 3CR sources. The allwise query delivers point source photometry in the 4 WISE bands ($W1/W2/W3/W4$ at $3.4/4.6/12/22~\micron$) and also the point source photometry from the 2MASS catalogue for $J, H$ and $K$ filters at $1.235, 1.662$ and $2.159~\micron$. 

Among the low-redshift sample six sources are extended, therefore PSF photometry was replaced by extended apertures\footnote{http://wise2.ipac.caltech.edu/staff/jarrett/wise/ext\_src.html} for 3C\,31, 3C\,35, 3C\,98, 3C\, 120, 3C\,236 and 3C\,390. 2MASS photometry for extended sources\footnote{http://tdc-www.harvard.edu/catalogs/tmxsc.html} was delivered by querying {\it fp\_xsc} catalogue with the IDL {\it query\_irsa\_cat} routine.

\subsection{Visible wavelengths}
\label{sec_data_optical}

A query on the SDSS catalogue (V/139/sdds9) with the IDL {\it query\_vizier} routine was performed. As not all sources were observed in SDSS, we complete the SEDs with data from \cite{Laing83} and \cite{Veron-Cetty10}. From the Hubble Space Telescope (HST) snapshot survey the host contribution in the visible was estimated by taking encircled energy diagrams (EEDs) from \cite{Lehnert99}. Emission line data for [\ion{O}{2}] and [\ion{O}{3}] were collected from \cite{Grimes04} and \cite{Jackson97}. For the medium-redshift sample [\ion{O}{3}] was measured for only 5 objects (3C\,207, 3C\,254, 3C\,263, 3C\,265 and 3C\,334).

\subsection{Radio wavelengths}
\label{sec_data_radio}

At radio wavelengths the data were collected from NED, which gives reference to the
following papers: 
%
\cite{Pilkington65,Pauliny-Toth66,Gower67,Aslanian68,Kellermann69,Colla70,Stull71,Colla72,Kellermann73,Fanti74,Laing80,Kuehr81,Large81,Geldzahler83,Ficarra85,Baldwin85,Hales88,Hales90,Becker91,Gregory91,Hales91,Becker95,Waldram96,Wiren92,Hales93,Gear94,Hales95,Griffith95,Klein96,Rengelink97,Condon98,Bennett03,Gilbert04,Mack05,Kassim07,Cohen07,Mantovani09,Wright09,Chen09,Chynoweth09,Jenness10,Agudo10,Richards11,Gold11,Algaba11,Lister11}

\section{Spectral Energy Distributions}
Figures~\ref{fig_seds_fsq1} to \ref{fig_seds_nlrg_rejected} show the
rest-frame SEDs for the sample at $0.5 < z < 1$,
separated for the AGN types with flat and steep radio spectrum (FSQ
and SSQ), HERGs with strong, medium and faint MIR emission, one
LERG and the 2 sources omitted from the analysis. 

Figures~\ref{fig_seds_low_fsq} to \ref{fig_seds_low_lerg2} show the
rest frame SEDs, separated for the FSQ and SSQ sources, BLRGs, HERGs
and LERGs, that are seen at redshifts $z < 0.5$.  The striking feature
of the Herschel PACS/SPIRE data is that they nicely bridge the former
gap between the radio and MIR SEDs. Also the WISE data points expand
the previous ISOCAM and Spitzer IRS/MIPS24 SED coverage. A
steep radio spectrum source is roughly constant in a $\nu F_{\nu}$
diagram, while a flat radio source rises towards shorter
wavelengths. For the host galaxy a synthetic stellar population from
\cite{Bruzual03} is used. The MIR emission is fitted with models for
clumpy tori from \cite{Hoenig10}. For the FIR a modified blackbody
(Eq.  \ref{equ_mbb}) with emissivity index $\beta = 1.5$ is used.

\subsection{SEDs at medium-redshifts}
\label{sec_med_sed_descrtiption}

\begin{figure*}
\centering
  \includegraphics[width=0.9\linewidth]{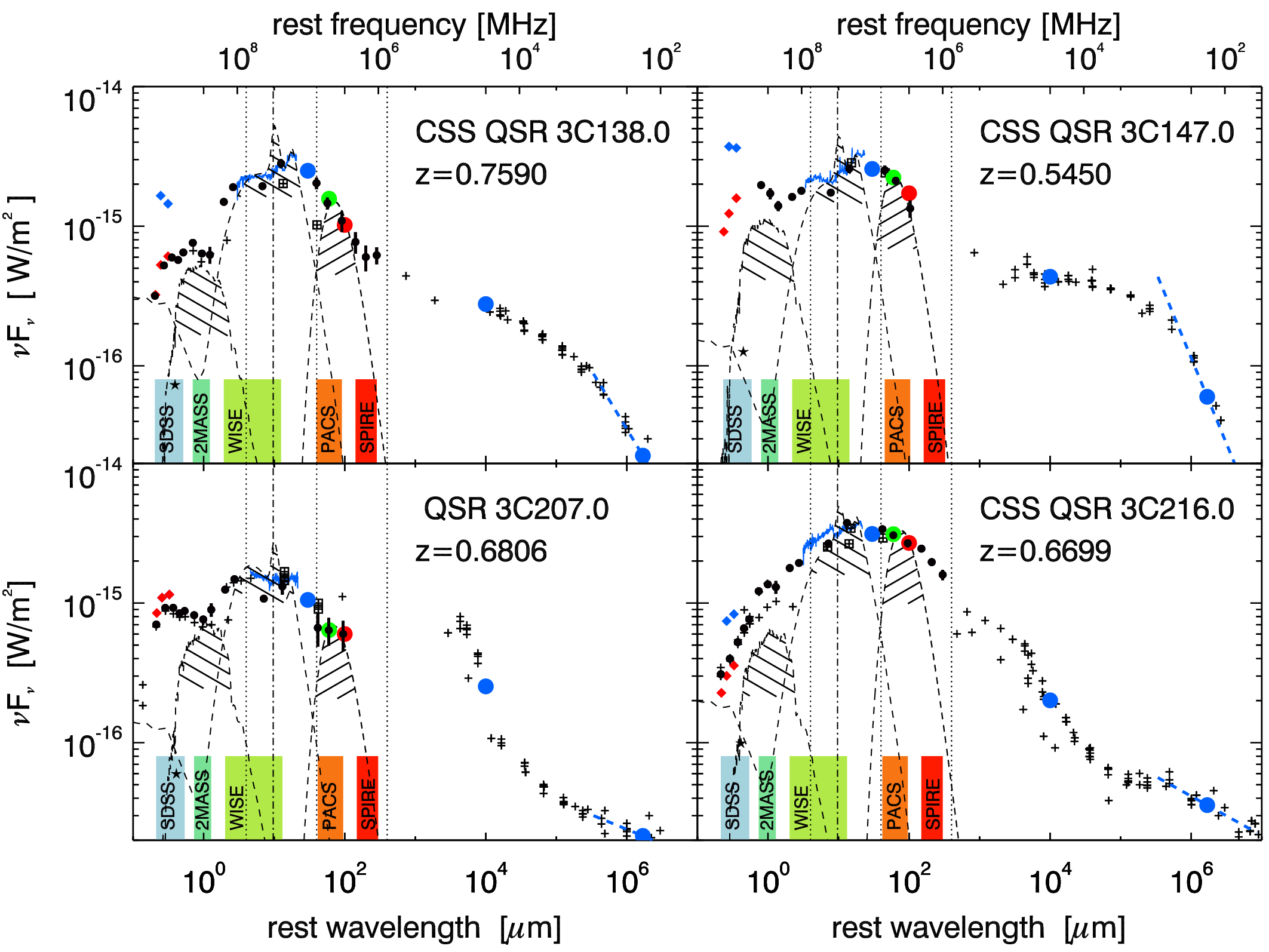}
  \caption{\label{fig_seds_fsq1} Spectral energy distributions of 4 flat
    spectrum quasars (FSQs) at $0.5 < z < 1$, i.e. quasars where the
    GHz respectively cm spectrum rises towards shorter wavelengths in
    $\nu F_{\nu}$ scaling. Filled black circles with error bars denote
    detections, 3$\sigma$ upper limits are marked by arrows. The
    Herschel PACS and SPIRE band ranges are shadowed in red, the 2MASS
    and WISE ranges in green, the optical (SDSS) range in blue. Red
    and blue diamonds are optical photometry values from
    \cite{Veron-Cetty10} and from \cite{Laing83}), respectively. ``+''
    symbols are detections (with arrows: upper limits) collected via
    NED. Disentangled host flux from \cite{Lehnert99} is shown with a star 
    symbol. Black open squares mark photometry with Spitzer/MIPS at
    $24~\micron$ or Spitzer/IRS; IRS spectra are plotted as blue lines
    and the position of the $9.7~\micron$ silicate absorption is
    indicated by the black vertical dash--dotted line. SCUBA
    $450$/$850~\micron$ and IRAM $1.2$\,mm data points by \cite{Haas04}
    are marked with black dots. Large blue dots mark median data
    points at 30\,GHz and 178\,MHz restframe. Big blue, green and red
    dots at IR-wavelengths mark interpolated flux levels at $30$, $60$
    and $100~\micron$, respectively. Models of the host galaxy, the AGN
    heated warm dust and the SF heated cool dust are shown as dashed
    lines.}
\end{figure*}
\addtocounter{figure}{-1}
\begin{figure*}
\centering
  \caption{\label{fig_seds_fsq2}continued}
\end{figure*}

\begin{figure*}
\centering
  \caption{\label{fig_seds_ssq} Spectral energy distributions of 6
    steep spectrum quasars (SSQs) at $0.5 < z < 1$, with strong
    optical AGN continuum and one BLRG without strong optical AGN
    continuum. Notation as in Fig.\ref{fig_seds_fsq1}.  }
\end{figure*}

\begin{figure*}
\centering
  \caption{\label{fig_seds_herg_mir_strong} Spectral energy
    distributions of high excitation radio galaxies (HERGs) at $0.5 <
    z < 1$, with bright MIR emission, up to a factor ten above the
    host galaxy level. Note the deep $9.7~\micron$ silicate absorption
    in the CSS 3C\,49 and in 3C\,55, 3C\,226 and 3C\,343.  Notation as
    in Fig.\ref{fig_seds_fsq1}.  }
\end{figure*}

\begin{figure*}
\centering
  \caption{\label{fig_seds_herg_mir_medium} Spectral energy
    distributions of high excitation radio galaxies (HERGs) at at $0.5
    < z < 1$, with medium MIR emission, reaching at about $30~\micron$
    the host galaxy level. Note the valley of low $4-10~\micron$
    emission in most sources. 3C\,330 and 3C\,441 are the only sources
    with successfully measured $9.7~\micron$ silicate absorption.
    Notation as in Fig.\ref{fig_seds_fsq1}. }
\end{figure*}

\begin{figure*}
\centering
  \caption{\label{fig_seds_herg_mir_weak} Spectral energy
    distributions of HERGs at $0.5 < z < 1$, with weak MIR
    emission. All sources have a low $4-10~\micron$ emission and barely
    reach at about $30~\micron$ the host galaxy level.  Notation as in
    Fig.\ref{fig_seds_fsq1}. }
\end{figure*}

\begin{figure*}
\centering
  \caption{\label{fig_seds_lerg} Spectral energy distributions of the
    only LERG in our sample at $0.5 < z < 1$. Note the low MIR flux
    compared to the host flux.  Notation as in
    Fig.\ref{fig_seds_fsq1}. }
\end{figure*}

\begin{figure*}
\centering
  \caption{\label{fig_seds_nlrg_rejected} Spectral energy
    distributions of two HERGs at $0.5 < z < 1$, which have been
    excluded from the analysis. 3C\,175.1 has too few data points and
    3C\,220.3 shows excess FIR-submm emission due to a gravitationally
    lensed submillimetre galaxy at z=2.2 (Haas et al. 2014).  Notation
    as in Fig.\ref{fig_seds_fsq1}. }
\end{figure*}

The quasar SEDs differ in their radio properties.  7 FSQs show a rise
in their GHz spectra (e.g. 3C\,207, Fig.~\ref{fig_seds_fsq1}), while
the 7 SSQs have GHz spectra which are constant in $\nu F_{\nu}$
(e.g. 3C\,175, Fig.~\ref{fig_seds_ssq}). A strong curvature is found
in the MHz to GHz spectra of some CSS quasars and radio galaxies,
(e.g. 3C\,147 in Fig~\ref{fig_seds_fsq1}, 3C\,343 in
Fig.~\ref{fig_seds_herg_mir_strong} and 3C\,343.1 in
Fig.~\ref{fig_seds_herg_mir_medium}); some CSS with modest
curvature are e.g. 3C\,196 in Fig~\ref{fig_seds_ssq}, 3C\,49 in
Fig.~\ref{fig_seds_herg_mir_strong} and 3C\,455 in
Fig.~\ref{fig_seds_herg_mir_medium}.  

We group 3C\,147 in the FSQs because of its SED rise between 90\,GHz
and 230\,GHz (\citealt{Steppe95}). The CSS 3C\,455 and 3C\, 343 are
sometimes classified as QSRs but they have neither a prominent 5\,GHz
core nor broad emission lines and therefore are treated here as HERGs
(Figs.~\ref{fig_seds_herg_mir_strong},~\ref{fig_seds_herg_mir_medium}). 

The SSQs show a $1.5$\,dex thermal bump in MIR--FIR
(Fig.~\ref{fig_seds_ssq}). However, the FSQs show a $\lesssim 0.5$\,dex
MIR-FIR emission bump above the extrapolated rising  GHz spectrum. The
FSQs  most likely have a strong synchrotron contribution to their IR
emission. 

At optical wavelengths the quasars (FSQ and SSQ) show a strong
power-law component rising towards shorter wavelengths. This component
and the hot dust emission at about $1~\micron$ outshine the host
galaxy. To estimate the host contribution in the SEDs we include the
disentangled host galaxy magnitude from HST imaging by
\cite{Lehnert99} to guide a fit for the host galaxy. 

Similarly to the SSQs, the HERGs (except 3C\,268.1, Fig~\ref{fig_seds_herg_mir_weak}) show a clear MIR--FIR emission
component above the extrapolation of the radio spectrum to shorter
wavelengths. However, there is a large diversity in the
MIR. Figures~\ref{fig_seds_herg_mir_strong} through
~\ref{fig_seds_herg_mir_weak} show the HERGs with strong, medium
and weak MIR emission relative to the host galaxy. 

While in both SSQs and HERGs the MIR SED is well determined, the
detection rate in the FIR (at rest frame $60$-$100~\micron$) is 17 detections out of a sample of 28 sources. The
sources with FIR detections are also bright in the MIR, and the SED
declines longward of $40~\micron$. Examples are 3C\,147 in
Fig.~\ref{fig_seds_fsq1}, 3C\,263 in Fig.~\ref{fig_seds_ssq}, and
3C\,226 in Fig.~\ref{fig_seds_herg_mir_strong}. The only exception
with a good detected FIR plateau beyond $40~\micron$ is 3C\,55 in
Fig.~\ref{fig_seds_herg_mir_strong}. The MIR-bright sources with FIR
upper limits mostly show the SED decline longward of $40~\micron$,
e.g., 3C\,254 in Fig.~\ref{fig_seds_ssq}, 3C\,22 in
Fig.~\ref{fig_seds_ssq}, and 3C\,280 in
Fig.~\ref{fig_seds_herg_mir_strong}. The remaining sources with FIR
upper limits often allow for an SED plateau beyond $40~\micron$
(Figs.~\ref{fig_seds_herg_mir_medium} and
~\ref{fig_seds_herg_mir_weak}).

The SED of the LERG 3C\,427.1 shows faint MIR (and FIR) emission
(Fig.~\ref{fig_seds_lerg}), corroborating the idea that LERGs are AGN with low
accretion activity (\citealt{Ogle06}). 

\subsection{SEDs at low-redshift}
\label{sec_low_sed_descrtiption}

\begin{figure*}
\centering
  \caption{\label{fig_seds_low_fsq} Spectral energy distributions of
    the flat spectrum sources at $z < 0.5$, three BLRGs and one
    quasar.  Notation as in Fig.\ref{fig_seds_fsq1}.  }
\end{figure*}

\begin{figure*}
\centering
  \caption{\label{fig_seds_low_ssq} Spectral energy distributions of
    steep spectrum quasars at $z < 0.5$.  Notation as in
    Fig.\ref{fig_seds_fsq1}.  }
\end{figure*}

\begin{figure*}
\centering
  \caption{\label{fig_seds_low_blrg} Spectral energy distributions of
    steep spectrum BLRGs at $z < 0.5$.  Notation as in
    Fig.\ref{fig_seds_fsq1}.  }
\end{figure*}

\begin{figure*}
\centering
 \caption{\label{fig_seds_low_nlrg1} Spectral energy distributions of
   high excitation narrow line radio galaxies with strong MIR emission
   at $z < 0.5$.  Notation as in Fig.\ref{fig_seds_fsq1}.  }
\end{figure*}

\begin{figure*}
\centering

  \caption{\label{fig_seds_low_nlrg2} Spectral energy distributions of
    high excitation narrow line radio galaxies with medium MIR
    emission at $z < 0.5$.  Notation as in Fig.\ref{fig_seds_fsq1}.  }
\end{figure*}

\begin{figure*}
\centering
  \caption{\label{fig_seds_low_nlrg3} Spectral energy distributions of
    high excitation narrow line radio galaxies with weak MIR emission.
    Notation as in Fig.\ref{fig_seds_fsq1}.  }
\end{figure*}

\begin{figure*}
 \centering
  \caption{\label{fig_seds_low_lerg} Spectral energy distributions of
    low excitation narrow line radio galaxies (LERGS) at $z < 0.5$.
    Notation as in Fig.\ref{fig_seds_fsq1}.  }
\end{figure*}

\begin{figure*}
\centering
  \caption{\label{fig_seds_low_lerg2} Spectral energy distributions of
    LERGS, with very low $L_{\rm {MIR}}/L_{\rm {Host}}$ at $z < 0.5$.
    Notation as in Fig.\ref{fig_seds_fsq1}.  }
\end{figure*}

The quasar/BLRG SEDs differ in their radio properties. Four flat spectrum
sources show a rise in their GHz spectra
(Fig.~\ref{fig_seds_low_fsq}), while the 10 steep spectrum sources
have GHz spectra which are constant in $\nu F_{\nu}$
(Figs.~\ref{fig_seds_low_ssq} and \ref{fig_seds_low_blrg}). A strong
curvature is found in the MHz to GHz spectra of two CSS quasars and
BLRGs, (3C\,48 in Fig~\ref{fig_seds_low_ssq},  and 3C\,268.3 in
Fig.~\ref{fig_seds_low_blrg}).  

The steep spectrum sources show a clear MIR--FIR emission component
above the extrapolation of the radio spectrum to shorter wavelengths
(Fig.~\ref{fig_seds_low_ssq}). In contrast, the flat spectrum sources show
a relatively modest MIR-FIR emission bump above the extrapolated
rising  GHz spectrum. They  most likely have a strong synchrotron
contribution to their IR SED. At optical wavelengths the quasars (2 of the 4 FSQs
and 3 of the 4 SSQs) show a strong power-law component rising towards shorter
wavelengths.  

Similarly to the SSQs, the HERGs show a clear MIR-FIR emission
component above the extrapolation of the radio spectrum to shorter
wavelengths. However, there is a large diversity in the
MIR. Figures~\ref{fig_seds_low_nlrg1} to \ref{fig_seds_low_nlrg3} show
the HERGs with strong, medium and weak MIR emission relative to the host galaxy. 

While in both SSQs and HERGs the MIR SED is well determined, the
detection rate in the FIR (at rest frame $60$--$100~\micron$) is
38/48. For the HERGs with strong and medium MIR emission the SED
declines longward of about $30$--$40~\micron$. 

The SEDs of the HERGs with weak MIR emission and of the LERGs show
faint MIR emission but mostly a rise towards the FIR
(Figs.~\ref{fig_seds_low_nlrg3} to \ref{fig_seds_low_lerg2}).

\subsection{Median SEDs of Quasars and Radio-Galaxies}
\label{sec_sed_median}

The SEDs of all sources were scaled to $\nu L_{\nu}$ with their
luminosity distance $D_{\rm L}$ given in Tables \ref{3C_int_lst} and
\ref{3C_low_lst}. The median SEDs were built for the classes FSQ,
SSQ, BLRG, HERG and LERG for each redshift sample as given in Tables
\ref{table_med_templ_seds} and \ref{table_low_templ_seds}. The
individual SEDs were first normalized to their 178\,MHz rest frame flux density,
which is interpolated and tabulated in Tables \ref{3C_med_lum_mono}
and \ref{3C_low_lum_mono} and then scaled to the median luminosity of
the sub-sample at 178\,MHz. The scaled SEDs were combined in continuous
bins of 100 consecutive data points. In each bin the median
wavelength, luminosity and standard deviation in logarithmic space was
calculated and plotted in Figure \ref{fig_templates}. The templates are tabulated in Table~\ref{tab_templ_all}.

The 178\,MHz restframe flux normalization was chosen because orientation effects can be excluded. Even so, the radio-lobe power may be influenced by the environment of the 3C sources. As shown in Section \ref{sec_result_dust_vs_radio} there is a trend of the ratio of radio-to-MIR luminosities changes with redshift, which can be interpreted as a denser environment at earlier ages. Therefore separate templates are provided for a range of source types and redshift ranges.

\begin{table*}
\centering
\scriptsize
\caption{Median spectral templates for different galaxy types separated for low- ($z < 0.5$) and medium-redshift ($0.5 < z < 1$).\label{tab_templ_all}}
\begin{tabular}{ccccccccc}
\hline
&&&&&&&&\\
&HERG-low-z&HERG-med-z&LERG&BLRG&FSQ-low-z&FSQ-med-z&SSQ-low-z&SSQ-med-z\\
&&&&&&&&\\
$log(\lambda_{Rest})$ & $log(\nu L_{\nu}^{})$  & $log(\nu L_{\nu}^{}) $ & $log(\nu L_{\nu}^{}) $ & $log(\nu L_{\nu}^{}) $ & $log(\nu L_{\nu}^{}) $ & $log(\nu L_{\nu}^{}) $ & $log(\nu L_{\nu}^{})$ & $log(\nu L_{\nu}^{}) $ \\
&&&&&&&&\\
$[\micron]$ & $[L_{\odot}]$ &$[L_{\odot}]$ &$[L_{\odot}]$ &$[L_{\odot}]$ &$[L_{\odot}]$ &$[L_{\odot}]$ &$[L_{\odot}]$ &$[L_{\odot}]$ \\
&&&&&&&&\\
\hline
&&&&&&&&\\
-0.4 & 10.4 $\pm$ 0.7 &  10.7 $\pm$ 0.4 &  &               &               &               11.8 $\pm$ 0.3 &  &               12.0 $\pm$ 0.6 \\
-0.3 & 10.7 $\pm$ 0.7 &  11.0 $\pm$ 0.3 &  10.6 $\pm$ 1.0 &  10.7 $\pm$ 0.5 &  11.0 $\pm$ 0.4 &  11.8 $\pm$ 0.2 &  11.6 $\pm$ 1.0 &  12.0 $\pm$ 0.6 \\
-0.2 & 11.0 $\pm$ 0.7 &  11.1 $\pm$ 0.3 &  10.8 $\pm$ 0.9 &  10.9 $\pm$ 0.5 &  10.9 $\pm$ 0.3 &  11.8 $\pm$ 0.2 &  11.6 $\pm$ 1.0 &  12.0 $\pm$ 0.6 \\
-0.1 & 10.9 $\pm$ 0.7 &  11.1 $\pm$ 0.3 &  11.0 $\pm$ 0.9 &  11.0 $\pm$ 0.4 &  10.9 $\pm$ 0.3 &  11.8 $\pm$ 0.2 &  11.6 $\pm$ 1.1 &  12.0 $\pm$ 0.4 \\
-0.0 & 11.0 $\pm$ 0.6 &  11.1 $\pm$ 0.2 &  11.0 $\pm$ 0.9 &  11.0 $\pm$ 0.4 &  10.9 $\pm$ 0.3 &  11.8 $\pm$ 0.2 &  11.6 $\pm$ 1.2 &  12.0 $\pm$ 0.4 \\
 0.1 & 11.1 $\pm$ 0.5 &  11.1 $\pm$ 0.2 &  11.2 $\pm$ 0.9 &  11.0 $\pm$ 0.3 &  10.8 $\pm$ 0.4 &  11.8 $\pm$ 0.2 &  11.7 $\pm$ 1.2 &  12.0 $\pm$ 0.4 \\
 0.2 & 11.0 $\pm$ 0.5 &  11.1 $\pm$ 0.2 &  11.1 $\pm$ 0.9 &  11.1 $\pm$ 0.3 &  10.9 $\pm$ 0.5 &  11.8 $\pm$ 0.2 &  11.7 $\pm$ 1.2 &  12.0 $\pm$ 0.4 \\
 0.3 & 10.9 $\pm$ 0.5 &  11.1 $\pm$ 0.2 &  11.1 $\pm$ 1.0 &  11.1 $\pm$ 0.3 &  11.0 $\pm$ 0.4 &  11.8 $\pm$ 0.2 &  11.7 $\pm$ 1.0 &  12.0 $\pm$ 0.4 \\
 0.4 & 10.9 $\pm$ 0.5 &  11.2 $\pm$ 0.2 &  10.5 $\pm$ 1.2 &  11.1 $\pm$ 0.3 &  11.0 $\pm$ 0.5 &  11.8 $\pm$ 0.2 &  11.7 $\pm$ 1.0 &  12.0 $\pm$ 0.3 \\
 0.5 & 10.7 $\pm$ 0.5 &  11.3 $\pm$ 0.3 &  10.3 $\pm$ 1.3 &  11.1 $\pm$ 0.4 &  11.0 $\pm$ 0.4 &  11.9 $\pm$ 0.2 &  11.7 $\pm$ 1.0 &  12.2 $\pm$ 0.3 \\
 0.6 & 10.6 $\pm$ 0.5 &  11.4 $\pm$ 0.3 &   9.9 $\pm$ 1.3 &  11.1 $\pm$ 0.4 &  10.9 $\pm$ 0.4 &  11.9 $\pm$ 0.2 &  11.7 $\pm$ 0.2 &  12.0 $\pm$ 0.2 \\
 0.7 & 10.6 $\pm$ 0.5 &  11.3 $\pm$ 0.3 &   9.5 $\pm$ 0.9 &  11.0 $\pm$ 0.5 &  10.9 $\pm$ 0.4 &  11.9 $\pm$ 0.2 &  11.8 $\pm$ 0.2 &  12.2 $\pm$ 0.3 \\
 0.8 & 10.7 $\pm$ 0.6 &  11.5 $\pm$ 0.3 &   9.6 $\pm$ 0.9 &  11.3 $\pm$ 0.5 &  11.0 $\pm$ 0.4 &  12.0 $\pm$ 0.2 &  11.9 $\pm$ 0.2 &  12.1 $\pm$ 0.3 \\
 0.9 & 10.7 $\pm$ 0.7 &  11.4 $\pm$ 0.3 &   9.4 $\pm$ 0.8 &  11.0 $\pm$ 0.5 &  10.9 $\pm$ 0.4 &  12.0 $\pm$ 0.2 &  11.9 $\pm$ 1.6 &  12.0 $\pm$ 0.3 \\
 1.0 & 10.8 $\pm$ 0.7 &  11.3 $\pm$ 0.3 &   9.4 $\pm$ 0.7 &  11.3 $\pm$ 0.5 &  10.9 $\pm$ 0.4 &  12.0 $\pm$ 0.2 &  12.0 $\pm$ 1.5 &  12.1 $\pm$ 0.3 \\
 1.1 & 10.8 $\pm$ 0.7 &  11.3 $\pm$ 0.3 &   9.6 $\pm$ 0.7 &  11.5 $\pm$ 0.5 &  10.8 $\pm$ 0.5 &  12.1 $\pm$ 0.2 &  11.7 $\pm$ 0.2 &  12.1 $\pm$ 0.4 \\
 1.2 & 10.9 $\pm$ 0.7 &  11.4 $\pm$ 0.4 &   9.8 $\pm$ 0.6 &  11.5 $\pm$ 0.4 &  10.7 $\pm$ 0.5 &  12.1 $\pm$ 0.2 &  12.1 $\pm$ 0.2 &  12.2 $\pm$ 0.4 \\
 1.3 & 10.9 $\pm$ 0.7 &  11.4 $\pm$ 0.3 &   9.8 $\pm$ 0.5 &  11.5 $\pm$ 0.5 &  10.7 $\pm$ 0.5 &  12.0 $\pm$ 0.2 &  12.1 $\pm$ 0.3 &  12.4 $\pm$ 0.3 \\
 1.4 & 10.8 $\pm$ 0.7 &  11.3 $\pm$ 0.5 &   9.5 $\pm$ 0.5 &  11.4 $\pm$ 0.5 &  10.6 $\pm$ 0.6 &  12.0 $\pm$ 0.2 &  12.2 $\pm$ 0.4 &  12.5 $\pm$ 0.5 \\
 1.5 & 10.8 $\pm$ 0.7 &  11.3 $\pm$ 0.5 &  10.6 $\pm$ 0.8 &  11.3 $\pm$ 0.5 &  10.5 $\pm$ 0.7 &  11.9 $\pm$ 0.2 &  12.1 $\pm$ 0.4 &  12.5 $\pm$ 0.5 \\
 1.6 & 10.8 $\pm$ 0.7 &  11.3 $\pm$ 0.5 &  10.7 $\pm$ 0.9 &  11.3 $\pm$ 0.5 &  10.4 $\pm$ 0.8 &  11.9 $\pm$ 0.3 &  12.1 $\pm$ 0.4 &  12.2 $\pm$ 0.7 \\
 1.7 & 10.8 $\pm$ 0.7 &  11.2 $\pm$ 0.6 &  10.6 $\pm$ 0.9 &  11.3 $\pm$ 0.5 &  10.3 $\pm$ 0.8 &  11.8 $\pm$ 0.3 &  11.9 $\pm$ 0.6 &  11.6 $\pm$ 0.8 \\
 1.8 & 10.5 $\pm$ 0.7 &  11.0 $\pm$ 0.7 &  10.4 $\pm$ 0.9 &  11.2 $\pm$ 0.7 &  10.2 $\pm$ 0.8 &  11.8 $\pm$ 0.3 &  11.8 $\pm$ 0.9 &  11.4 $\pm$ 0.8 \\
 1.9 & 10.3 $\pm$ 0.8 &  11.0 $\pm$ 0.7 &  10.0 $\pm$ 0.9 &  11.0 $\pm$ 0.8 &  10.3 $\pm$ 0.8 &  11.7 $\pm$ 0.3 &  11.7 $\pm$ 1.0 &  11.3 $\pm$ 0.8 \\
 2.0 & 10.2 $\pm$ 0.9 &  10.9 $\pm$ 0.7 &   9.8 $\pm$ 0.9 &  10.7 $\pm$ 0.9 &  10.3 $\pm$ 0.8 &  11.7 $\pm$ 0.3 &  11.6 $\pm$ 1.0 &  10.9 $\pm$ 0.8 \\
 2.1 & 10.1 $\pm$ 1.0 &  10.8 $\pm$ 0.6 &   9.7 $\pm$ 0.9 &  10.5 $\pm$ 0.9 &  10.2 $\pm$ 0.7 &  11.6 $\pm$ 0.3 &  11.3 $\pm$ 1.0 &  10.9 $\pm$ 0.8 \\
 2.2 &  9.9 $\pm$ 1.0 &  10.8 $\pm$ 0.6 &   9.5 $\pm$ 1.0 &  10.1 $\pm$ 0.8 &  10.2 $\pm$ 0.7 &  11.6 $\pm$ 0.3 &  11.1 $\pm$ 1.0 &  10.9 $\pm$ 0.8 \\
 2.3 &  9.7 $\pm$ 1.0 &  10.8 $\pm$ 0.6 &   9.5 $\pm$ 0.9 &   9.9 $\pm$ 0.8 &  10.2 $\pm$ 0.7 &  11.6 $\pm$ 0.3 &  11.0 $\pm$ 1.0 &  10.9 $\pm$ 0.8 \\
 2.4 &  9.6 $\pm$ 0.9 &  10.8 $\pm$ 0.6 &   9.3 $\pm$ 0.9 &   9.8 $\pm$ 0.8 &  10.2 $\pm$ 0.7 &  11.6 $\pm$ 0.3 &  10.9 $\pm$ 1.0 &  10.9 $\pm$ 0.8 \\
 2.5 &  9.5 $\pm$ 0.9 &  10.8 $\pm$ 0.6 &   9.2 $\pm$ 0.9 &   9.8 $\pm$ 0.8 &  10.2 $\pm$ 0.6 &  11.6 $\pm$ 0.3 &  10.7 $\pm$ 1.0 &  10.9 $\pm$ 0.8 \\
 3.0 &  9.4 $\pm$ 0.9 &  10.6 $\pm$ 0.6 &   9.2 $\pm$ 0.9 &   9.8 $\pm$ 0.8 &  10.1 $\pm$ 0.6 &  11.6 $\pm$ 0.3 &  10.3 $\pm$ 1.0 &  10.9 $\pm$ 0.8 \\
 3.4 &  9.4 $\pm$ 0.9 &  10.6 $\pm$ 0.6 &   9.2 $\pm$ 0.9 &   9.8 $\pm$ 0.8 &  10.0 $\pm$ 0.5 &  11.4 $\pm$ 0.2 &  10.3 $\pm$ 1.0 &  10.8 $\pm$ 0.8 \\
 3.8 &  9.3 $\pm$ 0.7 &  10.6 $\pm$ 0.6 &   9.2 $\pm$ 0.9 &   9.8 $\pm$ 0.8 &   9.8 $\pm$ 0.4 &  11.2 $\pm$ 0.3 &  10.3 $\pm$ 1.0 &  10.7 $\pm$ 0.8 \\
 4.2 &  9.1 $\pm$ 0.6 &  10.4 $\pm$ 0.6 &   9.2 $\pm$ 0.8 &   9.7 $\pm$ 0.7 &   9.6 $\pm$ 0.4 &  11.1 $\pm$ 0.3 &  10.3 $\pm$ 0.9 &  10.6 $\pm$ 0.7 \\
 4.6 &  9.0 $\pm$ 0.4 &  10.4 $\pm$ 0.3 &   9.1 $\pm$ 0.4 &   9.6 $\pm$ 0.3 &   9.3 $\pm$ 0.5 &  11.0 $\pm$ 0.3 &  10.2 $\pm$ 0.8 &  10.6 $\pm$ 0.2 \\
 5.0 &  9.0 $\pm$ 0.1 &  10.4 $\pm$ 0.3 &   9.1 $\pm$ 0.3 &   9.4 $\pm$ 0.1 &   8.8 $\pm$ 0.5 &  10.8 $\pm$ 0.3 &  &               10.5 $\pm$ 0.2 \\
 5.4 &  9.0 $\pm$ 0.2 &  10.3 $\pm$ 0.2 &   9.0 $\pm$ 0.1 &   9.4 $\pm$ 0.1 &   8.4 $\pm$ 0.4 &  10.5 $\pm$ 0.2 &  &               10.4 $\pm$ 0.1 \\
 5.8 &  8.9 $\pm$ 0.2 &  10.2 $\pm$ 0.2 &   8.9 $\pm$ 0.2 &   9.3 $\pm$ 0.1 &  &               10.4 $\pm$ 0.2 &  &               10.4 $\pm$ 0.1 \\
 6.2 &  8.8 $\pm$ 0.1 &  10.1 $\pm$ 0.1 &   8.8 $\pm$ 0.1 &   9.2 $\pm$ 0.2 &  &               10.2 $\pm$ 0.2 &  &               10.4 $\pm$ 0.1 \\
 6.6 &  8.7 $\pm$ 0.1 &  10.0 $\pm$ 0.2 &   8.7 $\pm$ 0.1 &   9.1 $\pm$ 0.1 &  &               &               &               10.3 $\pm$ 0.1 \\
&&&&&&&&\\
\hline
\end{tabular}
\end{table*}

While for HERGs, LERGs and also for BLRGs, the stellar component of
the SEDs is visible and can be fitted well, for the SSQs and FSQs the
strong power-law shaped AGN continuum in the optical and ultra-violet
has to be taken into account. Because that was not possible in a
consistent manner, the host galaxy fits and the derived stellar masses
have to be seen as upper limits for the quasars (Section
\ref{sec_sed_decomposition}). 

In the $\nu L_{\nu}$ scaling both quasar types, FSQ and SSQ, show a
flat, nearly identical distribution in the range from $0.1~\micron$
$\lesssim \lambda \lesssim 20~\micron$, justifying the assumption that
the two classes are intrinsically similar objects. At wavelengths
beyond $20~\micron$ the median SED of FSQs and SSQs diverges with a flux higher in the FSQs. This was interpreted
in the past as a jet component (\citealt{Cleary07}) which is
relativistically beamed towards us. Now with the new Herschel data
included, the jet enhancement can be traced to FIR wavelengths for the
FSQs, which also shows up in the non-thermal shape of the
SED. FIR-inferred SFR rates have
therefore only are upper limits for the FSQs (Section
\ref{sec_sed_decomposition}). 

The HERGs and LERGs median SEDs appear quite differently at IR wavelength. On average the HERGs are one dex more luminous in the MIR than the LERGs. The weak MIR activity of LERGs was interpreted as low accretion activity (\citealt{Ogle06}). For both types the MIR shows absorption features from silicate at $9.7~\micron$ absent in all quasar (FSQ, SSQ, BLRG) median SEDs. LERGs have a relatively weaker dust to starlight continuum ratio than HERGs. For LERGs the peak in the FIR is more distinguished and shifted to longer wavelengths, suggesting cooler dust compared to HERGs.


\begin{figure*}
\centering
\includegraphics[width=0.9\linewidth]{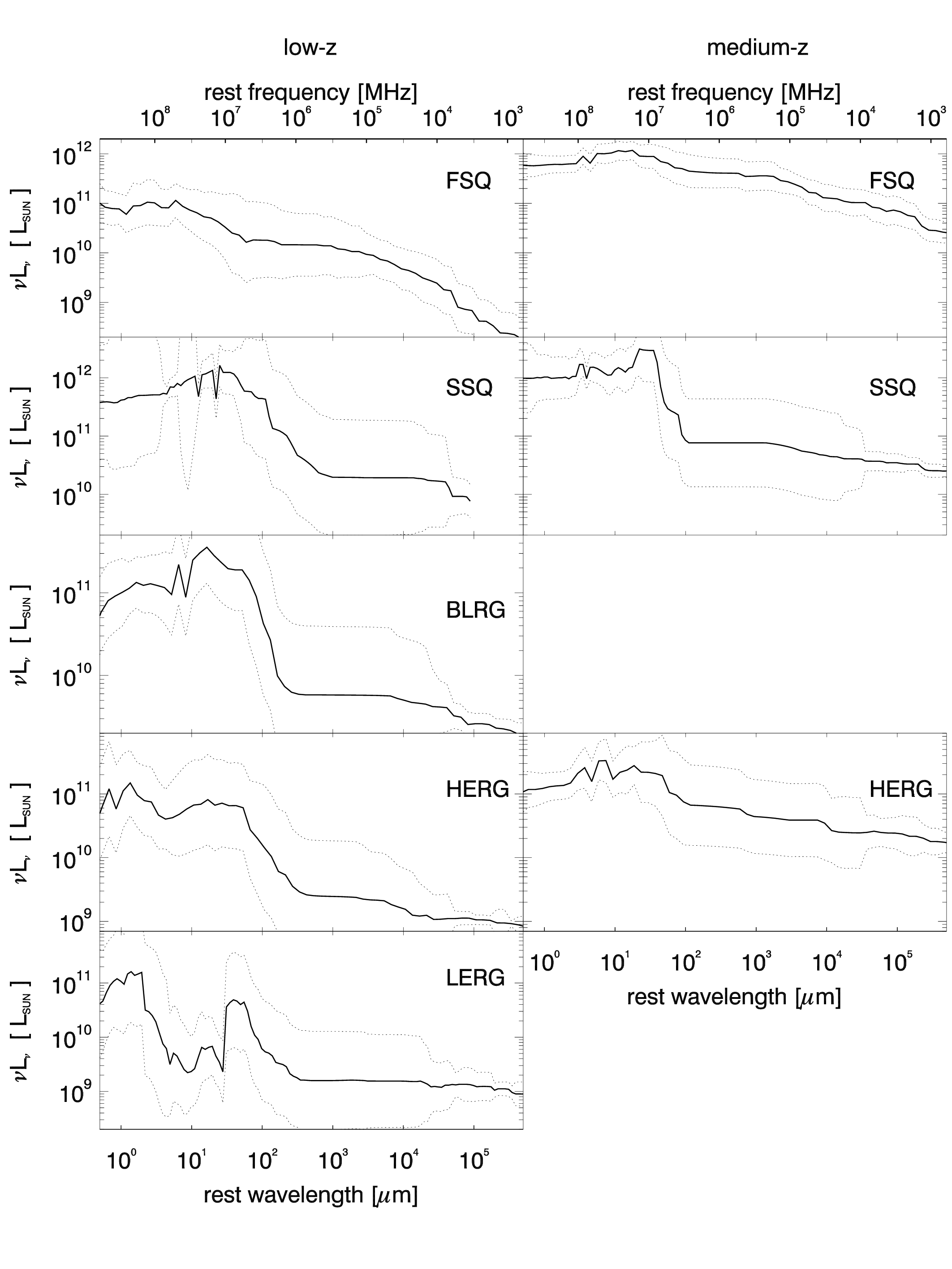}
\caption{Median of individual SEDs first normalized to their 178\,MHz rest frame flux and then scaled to the median luminosity of each  sub-sample (see Tables~\ref{table_med_templ_seds} and \ref{table_low_templ_seds}) at 178\,MHz, this normalization may be influenced by the source environment.\label{fig_templates}}
\end{figure*}



\subsection{Decomposition into Host, AGN Torus and Star formation}
\label{sec_sed_decomposition}

The components and structure of the galaxy (gas dust, stars, radio-jets and lobes) are reflected in the SEDs and can be disentagled from it. The stellar emission of the host galaxy peaks, depending on the stellar population, between NIR and UV wavelength (\citealt{Fioc97, Bruzual03, DeBreuck10})   

In the orientation-based unified scheme of powerful FRII radio
galaxies and quasars (\citealt{Barthel89, Antonucci93}) the optical
and UV emission of the central engine is blocked in some directions 
by anisotropically
distributed dust. The heating by the AGN causes the
warm dust emission to peak at restframe MIR ($10$--$40~\micron$) wavelength
(\citealt{Rowan-Robinson95}), which has been observed for most of the
3C sources (e.g., \citealt{Sieb04,Ogle06,Hardcastle09}). A torodial and
clumpy configuration in the so called dust torus is a widely accepted
hypothesis for the dust configuration (\citealt{Nenkova02, Hoenig06,Sieb15}).

The dust-enshrouded formation of stars causes the stellar light to be reprocessed by the dust. Corresponding to the cool temperature the re-emission peaks at $\sim~100\micron$ (\citealt{Schweitzer06, Netzer07, Veilleux09}). 

The aim of the analysis is to quantify host galaxy stellar mass and star formation rates in the environment of the
strong AGN emission, which can contribute at all wavelength
ranges. Also the question of the unification of radio galaxies and
quasars shall be answered at the FIR wavelengths, where the opacity is
low. From the SEDs are extracted:

{\it a) } $L_{\rm {Host}}$, the luminosity of the stars in
the host galaxy, by integration over fitted synthetic stellar
population models by \cite{Bruzual03} (libraries available from
Mariska Kriek \footnote{http://astro.berkeley.edu/\textasciitilde
  mariska/FAST\_Download.html}). From the synthetic stellar population
templates the luminosity of the stars in the host galaxy $L_{\rm
  {Host}}$ (Eq. \ref{equ_lhost}) was derived. With the inherent
mass-to-light ratio $\left(\frac{M}{L}\right)_{\rm {BC03}}$ of the
templates, stellar masses $M_{\star}$ can then be calculated
(Eq. \ref{equ_mstellar}). Values for both samples are given in Tables
\ref{3C_med_z_host_fit} and \ref{3C_low_z_host_fit}. The templates
were calculated with an exponentially declining star formation history
(with time scale $\upsilon$ [log yr]), metallicities $Z$ ranging
from sub- to super-solar, and a Chabrier IMF. Free parameters in the fitting routine were $\upsilon$, $Z$, and the age of the stellar population.
The synthesized flux-densities $F_{\lambda}^0$ were
attenuated for the extinction in the interstellar medium with the
dust-attenuation $k(\lambda)$ and $R_{\rm V} = 4.05$ 
(\citealt{Calzetti00}, see Eq.~\ref{equ_calzetti}). The dependence of
the derived stellar masses on the extinction coefficient $A_{\rm V}$ is
weak for a sample of early type galaxies (see
\citealt{Swindle11}). Therefore the median $A_{\rm V} = 0.1$ found for the
\cite{Swindle11} sample was applied here.

\begin{equation}
\label{equ_calzetti}
F_{\lambda}^{\rm {att}} = F_{\lambda}^0 \cdot 10^{-0.4 A_{\rm V} k(\lambda)/R_{\rm V}}
\end{equation}

\begin{equation}
\label{equ_lhost}
L_{\rm {Host}} = 4\pi D_{\rm L}^2\int{F_{\nu}^{\rm {BC03}}d\nu}
\end{equation}

\begin{equation}
\label{equ_mstellar}
M_{\star} = \left(\frac{M}{L}\right)_{\rm {BC03}} \times L_{\rm {Host}}
\end{equation}

{\it b) } $L_{\rm {AGN}}$, the luminosity of the AGN powered dust
(torus), by integration over fitted torus models by
\citealt{Hoenig10}. The MIR emission was fitted using a template
library\footnote{\url{http://www.sungrazer.org/CAT3D.html}}. Parameters
of the best fitting template with the derived total luminosity $L_{\rm
  {AGN}}$ (Eq. \ref{equ_lagn}) are given in Tables
\ref{3C_med_z_agn_fit} and \ref{3C_low_z_agn_fit}. Parameters used 
for the fitting
process were: The index $a$ of the power law for radial dust cloud
distribution, the number $N$ of clouds along the equatorial line of
sight, the half opening angle $\theta$, the optical depth $\tau$ and
the inclination to the observer $i$.
\begin{equation}
\label{equ_lagn}
L_{\rm {AGN}} = 4\pi D_{\rm L}^2 \int{F_{\nu}^{\rm {HK10}}d\nu}
\end{equation}

{\it c) } $L_{\rm {FIR}}$, the luminosity of cool FIR emitting dust,
by integration of a modified blackbody at 20--50~K ($\beta$ =
1.5), which is given by

\begin{equation}
\label{equ_mbb}
F_{\nu}^{\rm {MBB}} \propto \nu^{\beta} \ B_{\nu}(T_{\rm D})
\end{equation}

and

\begin{equation}
\label{equ_lfir}
L_{\rm {FIR}}  = 4\pi D_{\rm L}^2 \int{F_{\nu}^{\rm {MBB}}d\nu}
\end{equation}

The FIR emission can be attributed to dust heated by stars. The
integrated luminosity $L_{\rm {FIR}}$ was used to estimate the SFR by
applying Eq. \ref{equ_sfr} (taken from \citealt{Kennicutt98} Eq. (4)),
which is valid for starbursts with an age less than $10^8$
years. Values are given in Tables \ref{3C_med_z_fir_fit} and
\ref{3C_low_z_fir_fit}.

\begin{equation}
\label{equ_sfr}
{\rm SFR} \left[\frac{M_{\odot}}{{\rm yr}}\right] = 4.5\times10^{-44} L_{\rm {FIR}} \left[\frac{{\rm erg}}{{\rm s}}\right]
\end{equation}

{\it d) } Monochromatic luminosities at rest frame $\nu
  L_{\nu}^{30\,{\rm{GHz}}}$, $\nu L_{\nu}^{178\,{\rm{MHz}}}$ 
  were used to trace synchrotron
  contribution from radio jets and inclination effects in the IR at
  $\nu L_{\nu}^{30~\micron}$, $\nu L_{\nu}^{60~\micron}$, $\nu
  L_{\nu}^{100~\micron}$ from interpolated fluxes (Tables
  \ref{3C_med_lum_mono} and \ref{3C_low_lum_mono}).

  Present AGN torus models often require an ad hoc $T = 1300~{\rm K}$
  (\citealt{Leipski13,Podigachoski15a}) dust component in quasars to
  fit the near-infrared (NIR) bump around $3~\micron$. In addition, new
  AGN torus models \citep{Sieb15} invoke fluffy dust particles and are
  able to fit the AGN SEDs to longer wavelengths compared to the HK
  models, with SED peak beyond $\sim 80~\micron$. This increases the
  ambiguity of AGN-SF model fitting and star formation may be even
  lower than indicated by our analysis here.

\subsection{Bayesian SED fitting}
\label{sec_sed_fitting}

\begin{figure*}
\centering
\includegraphics[width=0.9\textwidth]{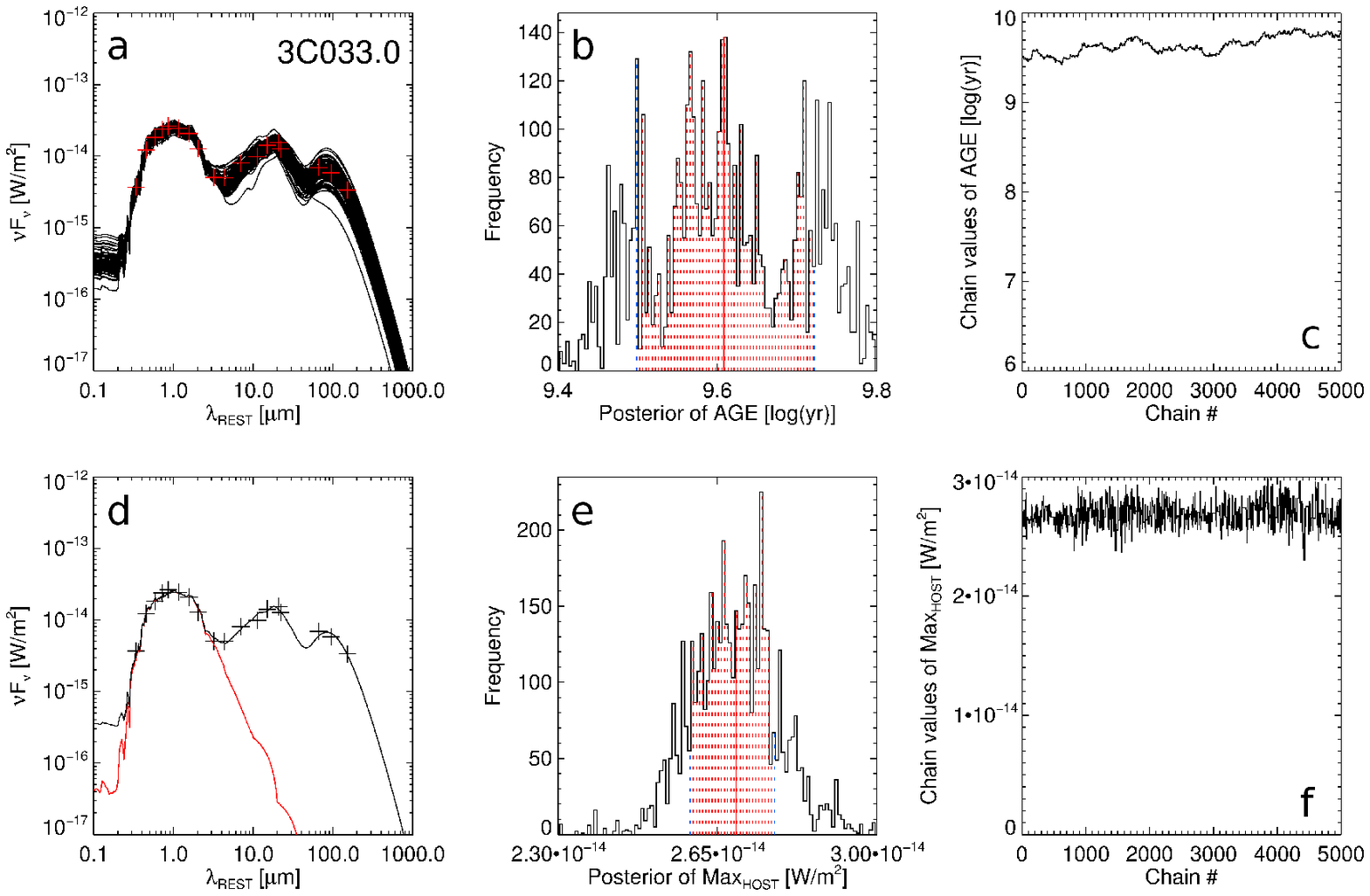}
\includegraphics[width=0.9\textwidth]{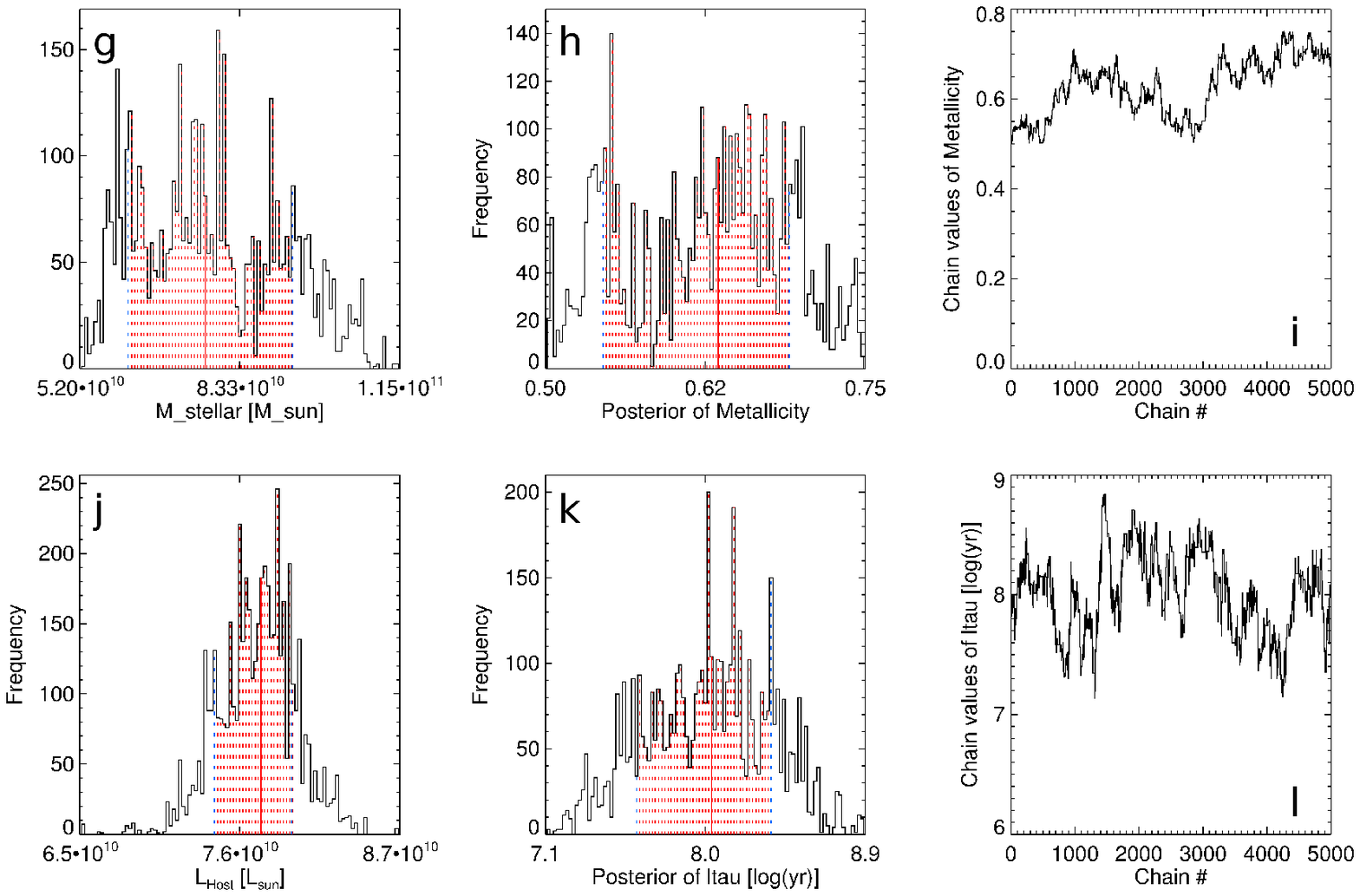}
\caption{Host fit: a) SED with black lines: range of considered models; d) SED with red line: best host template, black line: sum of best templates; b,e,g,h,j,k) distributions of host parameters and derived quantities, red solid line: median of distribution, red shaded area: 16\%--84\% interval of the total frequency; c,f,i,l) Monte-Carlo chains of host parameters.\label{fig_fit_host}}
\end{figure*}

\begin{figure*}
\centering
\caption{Torus fit: a) SED with red line: best torus template, black line: sum of best templates; b,d,e,g,h,j,k) distributions of torus parameters and derived quantities, red solid line: median of distribution, red shaded area: 16\%--84\% interval of the total frequency; c,f,i,l) Monte-Carlo chains of torus parameters.\label{fig_fit_torus}}
\end{figure*}

\begin{figure*}
\centering
\caption{FIR fit: a) SED with red line: best FIR template, black line: sum of best templates; b,d,e) distributions of FIR parameters and derived quantities, red solid line: median of distribution, red shaded area: 16\%--84\% interval of the total frequency; c,f) Monte-Carlo chains of FIR parameters.\label{fig_fit_fir}}
\end{figure*}

The fitting of all components was achieved by the application of a
Metropolis-Hastings algorithm under the investigation of the posterior
probability $Post(M|D)$ of the model $M$ fitting the data $D$, which
can be written after Baye's theorem (Eq. \ref{equ_bayes}) with
the prior of the model $Prior(M)$ and data $Prior(D)$ and the
likelihood of the data given the model $Like(D|M)$. 

\begin{equation}
\label{equ_bayes}
Post(M|D) = \frac{Like(D|M)Prior(M)}{Prior(D)}
\end{equation}

The prior $Prior(p)$ of a single model parameter $p$ in the range of
the maximum and minimum allowed values $p_{max}$ and $p_{min}$ is
given by the probability density of the uniform distribution (Eq.
\ref{equ_dunif}). 

\begin{equation}
\label{equ_dunif}
Prior(p) = \frac{1}{p_{max}-p_{min}}
\end{equation}

Then the prior $Prior(M)$ of the whole model can be written in
logarithmic space as Eq. \ref{equ_prior}. 

\begin{equation}
\label{equ_prior}
Prior(M) = \sum_i{\ln{Prior(p_i)}}
\end{equation}

The likelihood $Like(d|m)$ of single model point $m$ fitting a data
point $d$ with mean value $\mu$ and standard deviation $\sigma$ is
given by the probability density of the normal distribution (Eq.
\ref{equ_dnorm}). 

\begin{equation}
\label{equ_dnorm}
Like(d|m) = \frac{1}{\sqrt{2\pi}\sigma}e^{\frac{-(m-\mu)^2}{2\sigma^2}}
\end{equation}

The likelihood $Like(D|M)$ of the whole model $M$ fitting the data $D$
can then be written like Eq. \ref{equ_like}. 

\begin{equation}
\label{equ_like}
Like(D|M) = \sum_i{\ln{Like(d_i|m_i)}}
\end{equation}

With this nomenclature, a constant $Prior(D)$ and using the
logarithmic metric the posterior probability is calculated like
Eq. \ref{equ_post}. 

\begin{equation}
\label{equ_post}
Post(M|D) = Like(D|M)+Prior(M)
\end{equation}

Modelling parameters and ranges for host, torus and FIR templates are
given in Table \ref{table_params}. The Metropolis-Hastings Monte-Carlo
chain starts for each source with an individual set of scaling factors
for host-, torus- and FIR-template, while $Metal_{\rm {start}}=0.5$,
$\upsilon_{\rm {start}}=10^7$, $Age_{\rm {start}}=10^{9.5}$ for the
host $N_{\rm {start}} = 5.0$, $a_{\rm {start}}=-1$, $\theta_{\rm
  {start}}=32.5$ and $\tau_{\rm {start}}=35$ for the torus were
selected equally for all sources. The start value for inclination was
set to $5^{\circ}$ for type 1 sources and $45^{\circ}$ for type 2
sources. The start value for $T_{\rm {FIR}}$ was also selected
individually for each source.
The Metropolis-Hastings algorithm proceeds by randomizing the model
parameters $M_i$ of the preceding iteration with a proposal function
$M_{i+1}= func\_proposal(M_i)$ which was tuned to allow the chain
values to vary in suitable steps for each model parameter. Then the
posterior probability of the preceding model set $Post(M_i|D)$ was
compared and normalized to the new proposed model set
$Post(M_{i+1}|D)$ and the probability of an improvement was computed
via Eq. \ref{equ_probab} 

\begin{equation}
\label{equ_probab}
Prob(M_i|M_{i+1}|D) = e^{Post(M_{i+1}|D)-Post(M_i|D)}
\end{equation}

The computed value of $Prob(M_i|M_{i+1}|D)$ was compared to a
uniformally distributed random number between 0 and 1. If
$Prob(M_i|M_{i+1}|D)$ was greater than this random number the
proposed model set $M_{i+1}$ was included in the Monte-Carlo chain and
chosen as new start value for the next iteration step. This procedure
allows the chain to evolve to better models while also models which
don't seem to be an actual improvement retain a small chance to enter the
chain. By this behaviour the Metropolis-Hastings algorithm is able
to leave local maxima in the posterior space and search for the global
one. For each model 10000 chain values were calculated and the last
5000 iterations were used for the analysis via histograms for each
model parameter (see Figures \ref{fig_fit_host}, \ref{fig_fit_torus}
and \ref{fig_fit_fir}). 







\section{Results and discussion}
\label{chap_result}

\begin{figure*}[t!]
  \includegraphics[width=\linewidth]{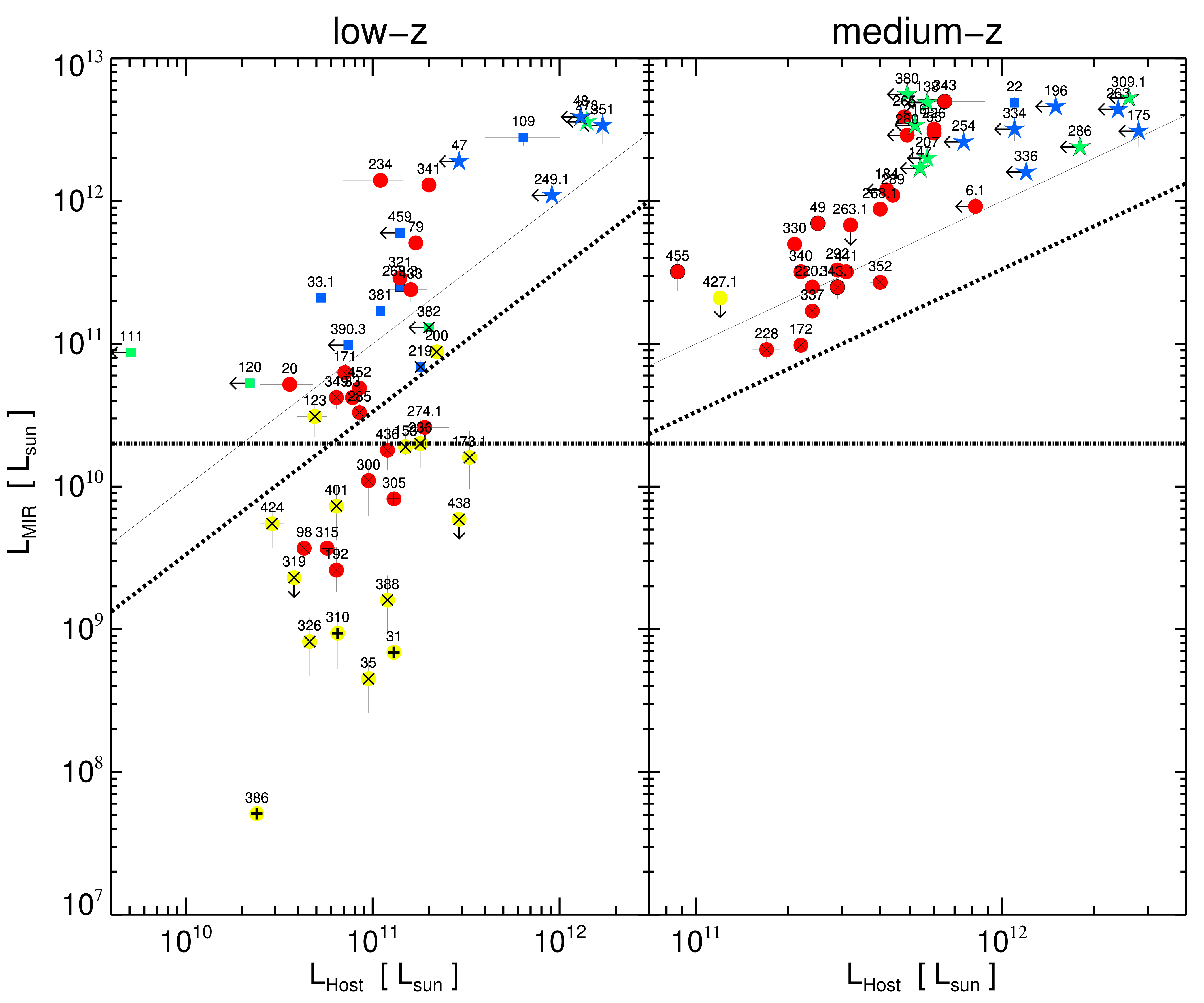}
  \caption{\label{fig_result_mir_weak_med} MIR vs host luminosities of
    the low ({\it {left}}) and medium ({\it {right}}) redshift
    samples. Circles indicate RGs (radio
    galaxies), stars indicate QSRs (quasars), squares BLRGs and arrows 
    3$\sigma$ upper limits.  SSQ (Steep spectrum QSRs) are in blue, FSQ
    (Flat Steep-spectrum QSRs) in green, HERG (high excitation radio
    galaxies) in red, LERG (low excitation radio galaxies) in yellow,
    and CSS (compact steep spectrum sources) in black, crosses
    indicate MIR-weak, and plus FRI (Fanaroff-Riley class I) sources,
    respectively.
    The dividing ratio of $L_{\rm {MIR}}/L_{\rm {Host}}=1$
    (solid line) separates the MIR-weak sources. A dividing ratio of $L_{\rm {MIR}}/L_{\rm {Host}}=1/3$ is denoted by the dotted line. The horizontal line
    at $2\times 10^{10} L_{\odot}$ corresponds to the absolute
    threshold defined by \cite{Ogle06}. }
\end{figure*}

\subsection{MIR-weak sources}
\label{sec_result_mir_weak}

Based on 15$~\micron$ luminosity measured in the Spitzer IRS spectra, Ogle
et al. (2006) defined MIR-weak sources by an absolute monochromatic
threshold, $\nu L_{\nu}^{15~\micron} < 8 \cdot 10^{43} {\rm erg/s}$ (roughly
corresponding to the integrated luminosity of the torus model fit of $L_{\rm {MIR}} = 2 \cdot 10^{10} L_{\odot}$). While
successfully identifying MIR-weak sources at low-redshift, 
potential analogs at higher redshift may be missed because they fall below the detection limit, and a more flexible
definition is desired -- also because $F_{\nu}^{15~\micron}$ is not
available for all of our sources.
The torus and host template fits are able to measure the entire integrated MIR luminosity as well as the host luminosity. Therefore we define MIR-weak sources relative to the host galaxy:  $L_{\rm MIR} / L_{\rm Host} < 1$ (Fig.~\ref{fig_result_mir_weak_med}).By this threshold all galaxies classified as LERGs are included in the MIR-weak definition, as well as some sources classified by their optical spectra as HERGs. Also two MIR-weak BLRGs, 3C\,219 and 3C\,382 are found according to this definition.

For 3C\,219 (see Fig.~\ref{fig_seds_low_blrg}) the MIR-weakness can be verified from the SED, where the fits of host and torus agree well with the observed data. For the flat-spectrum BLRG 3C\,382, the host luminosity cannot be independently estimated, and the model fit is most likely an over-estimate as contamination from the AGN could not be disentangled. A weaker ratio of $L_{\rm MIR} / L_{\rm Host} < \frac{1}{3}$ is also indicated in Fig.~\ref{fig_result_mir_weak_med} which would exclude the two BLRGs but also two LERGs and several HERGs from the MIR-weak definition.

The definition is motivated by the relation of black-hole and bulge masses (\citep{Haering04}), thus more massive galaxies can reach a larger accretion luminosity. To check the consistency of the inferred host masses and MIR luminosities, the black hole masses $M_{{\rm BH}}$ have been calculated with Eq. \ref{bh_bulge_relation} taken from \cite{Haering04} with stellar mass estimates $M_{\star}$ derived from the host luminosities $L_{{\rm Host}}$ as input (see Tables~\ref{3C_med_z_host_fit} and \ref{3C_low_z_host_fit}). We find black hole masses in the range of $\approx 10^6-10^9 M_{\odot}$ consistent with the range found for example by \cite{Tremaine02}

\begin{eqnarray}
  \label{bh_bulge_relation}
    \log(M_{{\rm BH}}/M_{\odot}) = (8.2 \pm 0.1)+(1.12 \pm 0.06)\\
    \nonumber
    \times \quad \log(M_{{\rm Bulge}}/10^{11}M_{\odot})
\end{eqnarray}

With the derived black hole masses $M_{{\rm BH}}$ we are able to calculate the limiting Eddington luminosity $L_{{\rm Edd}}$ as Eq.~\ref{eddington}. The ratio of AGN luminosity in the MIR and Eddington luminosity $L_{AGN}/L_{Edd}$ is given in Tables~\ref{3C_med_z_agn_fit} and \ref{3C_low_z_agn_fit}. The comparison shows an average ratio of a few percent with none of the sources exceeding the Eddington limit.

\begin{equation}
  \label{eddington}
  L_{{\rm Edd}} \approx 3.2 \times 10^4 \left(\frac{M_{{\rm BH}}}{M_{\odot}}\right) L_{\odot}
\end{equation}

The definition of MIR-weak sources relative to the hosts has the advantage that it is independent of absolute luminosities and thus may allow us to identify MIR-weak sources also at higher redshift. (In fact, we find MIR-weak sources which exceed Ogle et al.'s absolute luminosity limit by about a factor of ten). In the plots, MIR-weak sources are marked with a superposed "x".


Beyond $z > 0.5$, the 3C-sample contains 5 MIR-weak HERGs but 
only one LERG 3C\,427.1 ($L_{\rm {MIR}}$ upper limit, Fig.~\ref{fig_result_mir_weak_med}). 
The lack of LERGs raises the question whether (some) MIR-weak HERGS were misclassified
and actually belong to the LERG class. 
We checked the classification into low- and high-excitation emission
line sources based on the ratio of [\ion{O}{2}]/[\ion{O}{3}] $>$ 1 (LERG) and [\ion{O}{2}]/[\ion{O}{3}] $<$
1 (HERG), using the spectroscopic data from \cite{Jackson97} and
\cite{Grimes04}. 
Due to the shift of the [\ion{O}{3}] line from restframe
wavelength of 5007~\AA\, to infrared wavelengths only 7 sources of the
medium-redshift sample (3C\,207, 3C\,220.1, 3C\,254, 3C\,263, 3C\,265,
3C\,280 and 3C\,334) have measured [\ion{O}{3}] fluxes. The other [\ion{O}{3}]
fluxes given by \cite{Grimes04} have been extrapolated from 
\cite{Jackson97} [\ion{O}{2}] fluxes using an average HERG line ratio.
 Therefore misclassification among the MIR-weak
HERGs of the medium-redshift sample cannot be excluded.
 
For the low-redshift sample the coverage of tabulated emission line
data in [\ion{O}{2}] and [\ion{O}{3}] is better and consistent with
the HERG classifications (except for the
starburst galaxy 3C\,459), 
but the classification could only be confirmed for one LERG (3C\,236), the others have no emission line data available.

We find two MIR-weak  BLRGs (3C\,219 and 3C\,382);
they could be the broad-line counterparts of
the otherwise type-2 dominated MIR-weak class. 
This finding is remarkable because type-1 AGN 
are typically brighter in the MIR than type-2 AGN, 
and the type-1 hosts are more difficult to measure.
Nevertheless, the number ratio of the type-1 / type-2 MIR-weak is small and a large dust covering angle would be required 
to reach consistency with orientation based unification schemes.

In the orientation-based AGN unification, 
MIR-weak sources either have less dust, a dust torus with a small covering angle
or low accretion power. To distinguish between these scenarios is 
a particular challenge. MIR-weak sources are found among 
both HERGs and LERGs, indicating a potential smooth transition and
arguing against sharply distinguished, fundamentally different
mechanisms like "radiation dominated" vs. "advection dominated" 
accretion (\citealt{Ogle06}).

\subsection{Comparison of MIR to radio-lobe and [\ion{O}{3}] luminosities}
\label{subsec_result_mir_vs_radio_OIII}

Three luminosities in MIR, radio-lobe and [\ion{O}{3}] luminosities are expected to be
tracers of the intrinsic AGN accretion power and therefore should be
correlated. 
\begin{figure*}[t!]
  \includegraphics[width=\linewidth]{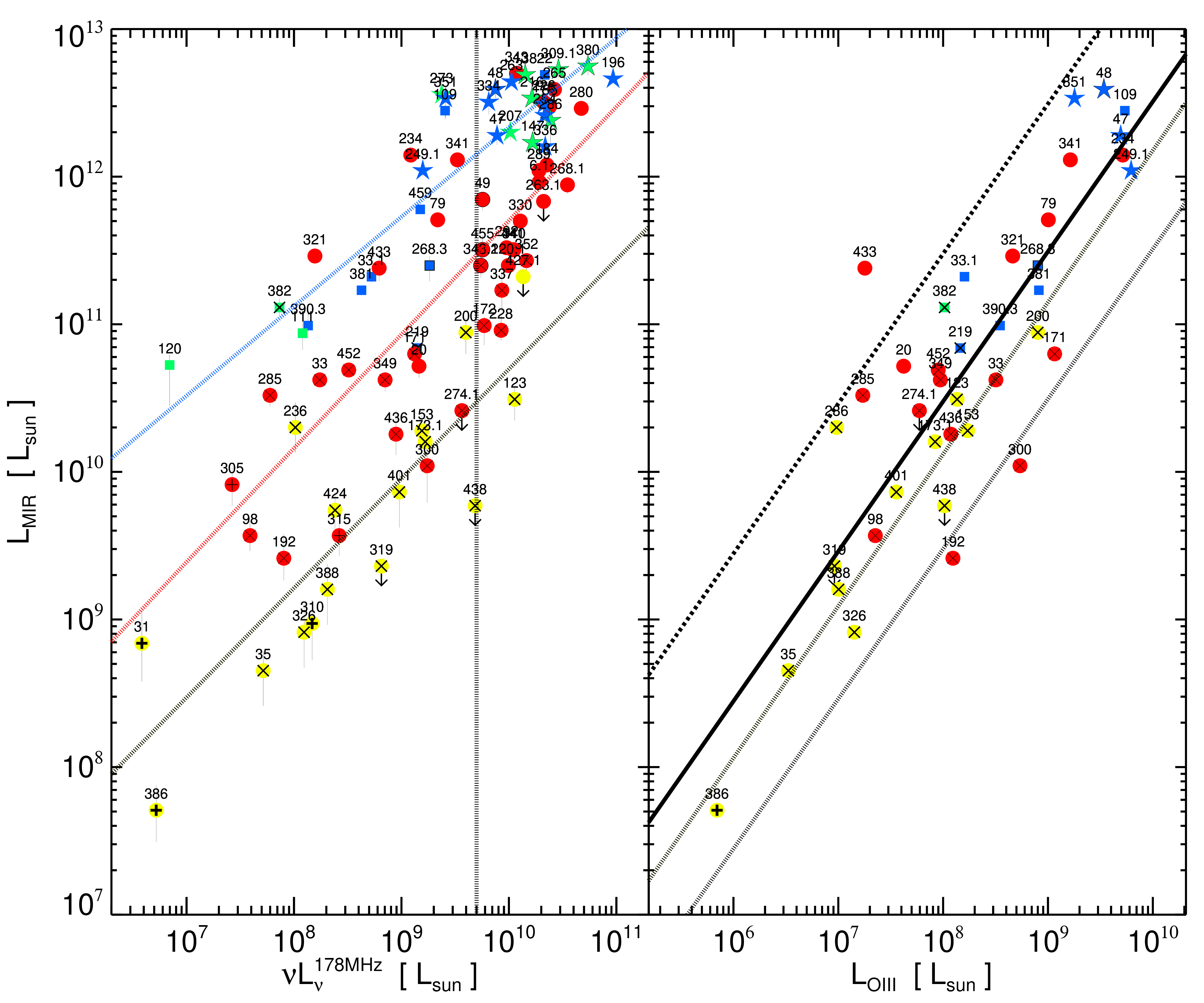}
  \caption{\label{fig_result_mir_radio} {\it left}: MIR luminosity versus radio luminosity. The vertical dashed line at $5\cdot10^9 L_{\odot}$ radio power roughly separates the medium from the low-redshift sample (exceptions are the low-$z$ sources 3C\,47, 3C\,48 and 3C\,123). The colored dotted lines denote fits to the SSQs (blue), HERGs (red) and LERGs (yellow).{\it right}: MIR luminosity versus [\ion{O}{3}] luminosity. Only observed [\ion{O}{3}] fluxes from \citep{Grimes04} are included. Black solid line denotes median relation of all plotted sources. Dashed and dotted black lines denote one dex range from median relation. Yellow dotted line denotes median relation for LERGs. Notation as in Fig.~\ref{fig_result_mir_weak_med}.}
\end{figure*}
Fig.~\ref{fig_result_mir_radio} (right) shows the ratio $L_{\rm {MIR}}/L_{\rm{O}\,III}$, which appears similar over four orders of magnitude for all three classes QSRs, HERGs and LERGs. The relation was fitted for all classes (see Eq. \ref{mir_oiii_slope_all}) and for the LERGs separately (see Eq. \ref{mir_oiii_slope_lerg}). There is a trend that LERGs have about a factor 3 higher $L_{\rm{O}\,III}$ compared to QSRs and HERGs of the same $L_{\rm {MIR}}$ (all LERGs except 3C\,236 lie on the right side of the median relationship in Fig.~\ref{fig_result_mir_radio},(right)). This trend can be explained by differences in the dust torus or intrinsic difference in the AGN-SED of high- and low-excitation sources. A smaller torus covering angle and less extinction of the inner NLR might be the cause. Also a central engine with a lower production rate of ionizing photons is thinkable. Best-fit relations are:


\begin{eqnarray}
\label{mir_oiii_slope_all}
\log(L_{\rm {MIR}}^{\rm {All}}/L_{\odot}) = (2.4 \pm 0.8)+(1.0 \pm 0.1) \\
\nonumber
\times \quad \log(L_{\rm{O}\,III}/L_{\odot})
\end{eqnarray}

\begin{eqnarray}
\label{mir_oiii_slope_lerg}
\log(L_{\rm {MIR}}^{\rm{LERG}}/L_{\odot}) = (1.9 \pm 0.7)+(1.0 \pm 0.1) \\ 
\nonumber
\times \quad \log(L_{\rm{O}\,III}/L_{\odot})
\end{eqnarray}

In Fig.~\ref{fig_result_mir_radio} (left) the monochromatic radio-lobe luminosity is plotted versus the MIR luminosity. The distributions for the different classes of QSRs, HERGs and LERGs show a clear overall trend.  The correlations for QSRs, HERGs and
LERGs are fitted separately (see Eq. \ref{mir_radio_slope_qsr} - \ref{mir_radio_slope_lerg}). At $\nu L_{\nu}^{178\,{\rm{MHz}}} = 5\cdot10^9 L_{\odot}$ the low from the medium-redshift sample are separated; there are only three low-$z$ exceptions (3C\,47, 3C\,48 and 3C\,123) exceeding that radio luminosity threshold.

\begin{eqnarray}
\label{mir_radio_slope_qsr}
\log(L_{\rm {MIR}}^{\rm{QSR}}/L_{\odot}) = (6.3 \pm 0.6)+(0.61 \pm 0.07) \\
\nonumber
\times \quad \log(L_{\nu}^{178\,{\rm{MHz}}}/L_{\odot})
\end{eqnarray}

\begin{eqnarray}
\label{mir_radio_slope_herg}
\log(L_{\rm {MIR}}^{\rm{HERG}}/L_{\odot}) = (4.0 \pm 0.9)+(0.8 \pm 0.1) \\
\nonumber
\times \quad \log(L_{\nu}^{178\,{\rm{MHz}}}/L_{\odot})
\end{eqnarray}

\begin{eqnarray}
\label{mir_radio_slope_lerg}
\log(L_{\rm {MIR}}^{\rm{LERG}}/L_{\odot}) = (3.3 \pm 1.1)+(0.8 \pm 0.1) \\
\nonumber
\times \quad \log(L_{\nu}^{178\,{\rm{MHz}}}/L_{\odot})
\end{eqnarray}

The HERGs and LERGs show a remarkably similar $L_{\rm MIR}/L_{\nu}^{178\,{\rm{MHz}}}$ slope, apart
from the offset, while the quasars show a flatter  $L_{\rm MIR}/L_{\nu}^{178\,{\rm{MHz}}}$ slope.
The torus models do not properly account for the hot
($\sim$1000\,K) dust emission in type-1 sources. This was
already noted by \cite{Leipski13} and \cite{Podigachoski15a} for the
high-z quasars.  
Thus the integrated luminosity $L_{\rm MIR}$ from the fitted torus models is underestimating the MIR luminosities of the bright quasars.
Independent of the slopes, 
the MIR/radio ratio of QSRs exceeds that of the HERGs by a factor of
5-10. 
This can
be explaind by orientation-dependent extinction even at MIR
wavelengths  (e.g. \citealt{Haas08,Leipski10, Podigachoski15b}).


Likewise, both LERGs and MIR-weak HERGs show a 
2--20 times weaker $L_{\rm {MIR}}/\nu L_{\nu}^{178\,{\rm{MHz}}}$ ratio
than the MIR-strong HERGs which is rather caused by decreased MIR than increased radio luminosity. Differences in the central engine can cause different AGN SEDs or low dust content may be the reason for MIR weakness; for example a binary black hole or differences in the black hole spin could lead to strong jet development on lower accretion rates.


\begin{figure*}

  \includegraphics[width=\linewidth]{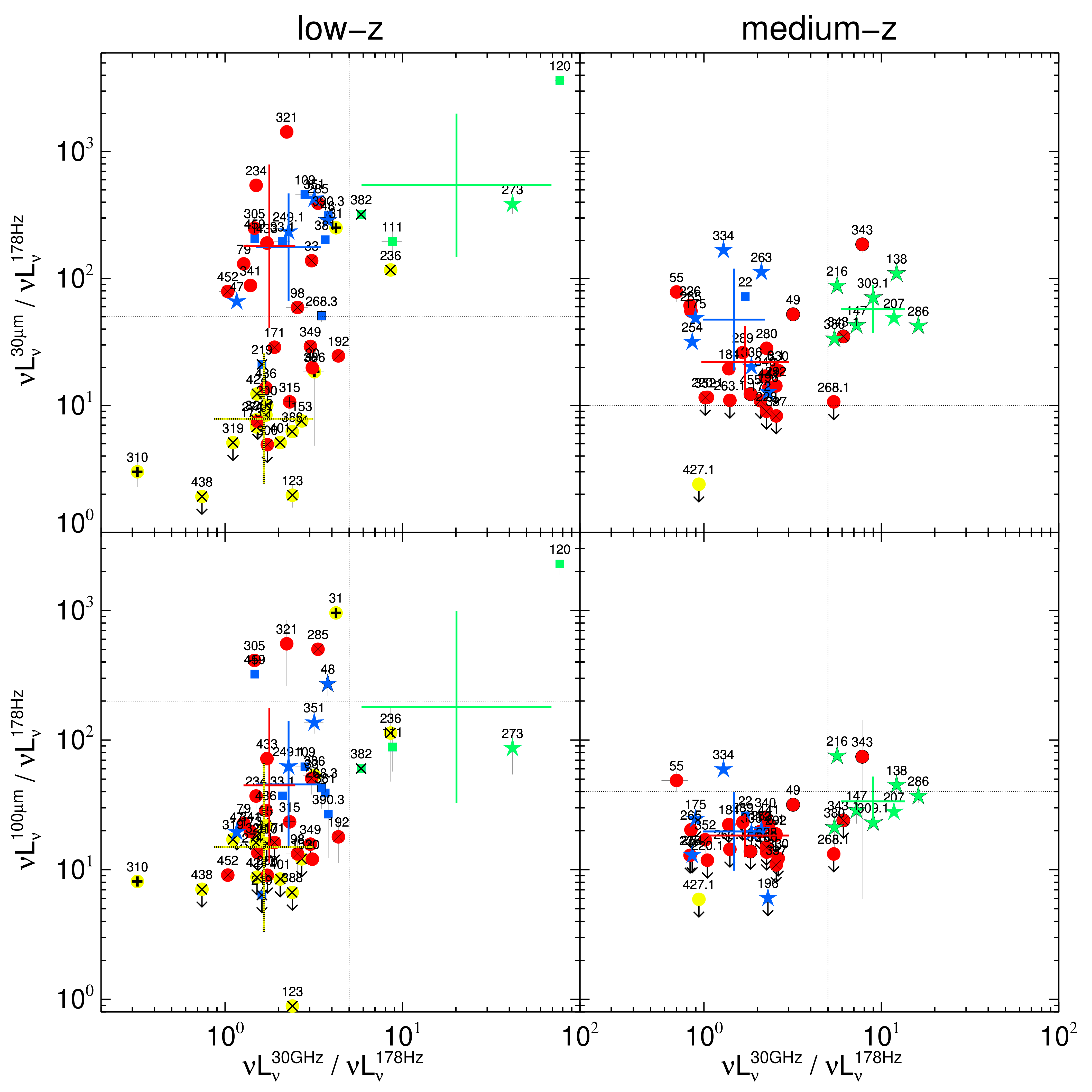}

  \caption{\label{fig_result_rdr030_med} IR and radio luminosity
    ratios for low ({\it left}) and medium ({\it right}) redshift
    samples.  $R_{\rm {DR}} = \nu F_{\nu}^{\rm {IR}}
/ \nu F_{\nu}^{178\,{\rm{MHz}}}$ at 30 and 100~$\micron$ are shown in the top
    and bottom panels respectively. Large crosses give the logarithmic mean and the
    dex range of the HERGs (red), LERGs (yellow), SSQs (blue) and all
    FSQs (green), CSS are excluded The vertical dotted line at $R_{\rm
      {GM}}=5$ separates FSQs from the other AGN types.  The
    horizontal dotted lines separate in the top panels (low-$z$: $R_{\rm
      {DR}}=50$, medium-z: $R_{\rm {DR}}=10$) LERGs and MIR-weak
    sources, and in the bottom panels (low-$z$: $R_{\rm {DR}}=200$,
    medium-z: $R_{\rm {DR}}=40$) FIR/radio excess sources.  Notation
    as in Fig.~\ref{fig_result_mir_weak_med}.}
\end{figure*}

\subsection{IR and radio luminosity ratios}
\label{sec_result_dust_vs_radio}

\subsubsection{Beamed IR contribution in FSQs}
\label{sec_result_beamed}

At medium redshifts the SSQs 3C\,334 and 3C\,336 (and also some HERGs) have such low 1.2\,mm fluxes observed (\citealt{Haas04}) that there is no room for a beamed synchrotron
component (with reasonable spectral slope $\alpha$) to contribute to
the MIR and FIR (Fig.~\ref{fig_seds_ssq}). This is comparable to radio-quiet
quasars as seen by \cite{Chini89}. The other 6 SSQs do not have 1.2\,mm 
measurements, but the same picture can be assumed for them. 

In contrast the 7 FSQs have a rising GHz spectrum and beamed
emission may contribute to the FIR and MIR
(Fig.~\ref{fig_seds_fsq1}). A strong FIR contribution is immediately
obvious for 3C\,138, 3C\,216 and 3C\,286. In the MIR, however, a sharp
bump at $\sim~20~\micron$ can be identified in  most
sources, except 3C\,207 and 3C\,216. Also the NIR 3--4~$\micron$ bump is
discernible  (e.g., 3C\,147). This suggests that the NIR--MIR SED is
dominated by dust emission and that any beamed contribution to the MIR
and NIR is weaker than in the FIR. 
The same picture can been seen at lower redshifts for the FSQs in
Fig.~\ref{fig_seds_low_fsq}, where the non-thermal contribution can be
traced to the FIR for 3C\,111, 3C\,120, 3C\, 273 and 3C\,382. The SSQs
and BLRGs plots show (Fig.~\ref{fig_seds_low_ssq} and
\ref{fig_seds_low_blrg}) that the FIR is dominated by dust emission.

Following the concept introduced by Meisenheimer et al. (2001), we
determined the dust-to-radio ratio $R_{\rm {DR}} = \nu F_{\nu}^{\rm {IR}}
/ \nu F_{\nu}^{178\,{\rm{MHz}}}$ at $30$ and $100~\micron$.  To quantify
the dependence of $L_{IR}$ at 30 and 100$~\micron$ on the radio slope,
we consider $R_{\rm {DR}}$ versus $R_{\rm {GM}} = \nu
L_{\nu}^{30\,{\rm{GHz}}}/\nu L_{nu}^{178\,{\rm{MHz}}}$
(Fig.~\ref{fig_result_rdr030_med}).  $R_{\rm {GM}}$ clearly separates
FSQs from SSQs, with a dividing ratio $R_{\rm {GM}} = 5$.
Fig.~\ref{fig_result_rdr030_med} shows on the vertical axes the
$R_{\rm {DR}}$ distributions at 30, and 100~$\micron$ for the different
AGN types.  The large crosses mark the averages (in log space) and the
dex range for the different AGN types:

\begin{itemize}

\item[$\bullet$]
yellow/black: LERGs, excluding FR\,I and 3C\,236 as outlier

\item[$\bullet$]
red: HERGs, excluding CSS, FR\,I and MIR-weak sources

\item[$\bullet$]
blue: SSQs, i.e. steep spectrum QSRs and BLRGs excluding CSS 

\item[$\bullet$]
green: FSQs, i.e. flat spectrum QSRs and BLRGs

\end{itemize}

\subsubsection{Cosmological evolution of radio activity}

$R_{\rm {DR}}^{30~\micron} = 10$ for the medium-redshift sample and $R_{\rm {DR}}^{30~\micron} = 50$ for the low-redshift sample can be used as a separator of MIR-weak and MIR-strong sources. $R_{\rm {DR}}^{100~\micron} = 200$ for the low- and $R_{\rm {DR}}^{100~\micron} = 40$ for the medium-redshift sample separate sources with an
exceptional high FIR luminosity like 3C\,321 or the CSS 3C\,48; the curved radio spectrum of the CSS leads to an even lower
178\,MHz flux.  
This check reveals that statistically $L_{\rm {FIR}}$ is higher by a factor
of 2-3 in the FSQs. In contrast the MIR luminosity is similar (except 3C\,120), which can be explained by a wavelength-independent constant
contribution of non-thermal radiation, which is more dominant in the
sources with weaker (about 3 times) IR emission. 

Fig.~\ref{fig_result_rdr030_med} shows a remarkable
$R_{\rm {DR}}^{30~\micron}$ and $R_{\rm {DR}}^{100~\micron}$ difference between the
low and medium redshift samples of SSQs and MIR-strong HERGs.
For the medium-$z$ sample, on average, $R_{\rm {DR}}$ is about a factor 5--10 lower
than for the low-$z$ sample. This indicates that at a given MIR AGN power
the radio lobes are much fainter in the local universe
compared to the earlier epoch. Thus, the lobe production via
working surface of the jet with the ambient medium is
less efficient.
Notably, $R_{\rm {DR}}$ of the local MIR-weak
HERGs matches that of the distant MIR-strong
HERGs and the local LERGs have the lowest $R_{\rm {DR}}$.
This may indicate that their circumgalactic medium is
denser, perhaps due to cluster environment
(further discussed in Sect.~\ref{sec_result_evolution}). 

\subsubsection{SSQ / MIR-strong HERG unification}
\label{sec_result_unification}

Both the radio-lobe luminosity $\nu L_{\nu}^{178\,{\rm{MHz}}}$ and the
dust luminosity $L_{\rm {FIR}}$ are assumed to be isotropic. In the
orientation-based unified scheme, their distributions should be
indistinguishable for steep spectrum quasars (SSQs) and
high-excitation radio galaxies (HERGs).  The same should hold for {\it
  ratios} of isotropic observables. If the distributions differ, then
either the observables are not isotropic or the sources have intrinsic
differences.  For the low-redshift sample the HERGs (with the MIR-weak
ones excluded) and SSQs sub-samples show a nearly perfect match in
logarithmic average and range (Fig.~\ref{fig_result_rdr030_med}). FRI
and CSS sources were excluded from the average for both classes.



For the medium-redshift sample the SSQs show a $R_{\rm {DR}}$ ratio at
$30~\micron$ that is on average a factor of two higher than the HERGs,
but the samples match in their average $R_{\rm {DR}}$ at $60~\micron$ and even
better at $100~\micron$.  This can be interpreted by a dust torus that is
optically thick at $30~\micron$ and emits isotropically at $100~\micron$.
Also the FSQs match with SSQs in the unification framework by
synchrotron contribution from beamed jet emission in FIR and MIR. The
match at $100~\micron$ may be favoured by the fact that the averages are
dominated by the match of the upper limits. Again MIR-weak and CSS
sources were excluded.

These results support the orientation-based unified scheme 
for MIR-strong HERGs and SSQs. 
However, the relationship to 
LERGs and MIR-weak HERGs might not be explained simply by orientation
effects, as discussed in Sect.~\ref{sec_result_evolution}.

\subsection{Star formation}
\label{sec_result_star_formation}

\begin{figure*}[t!]
  \includegraphics[width=\linewidth]{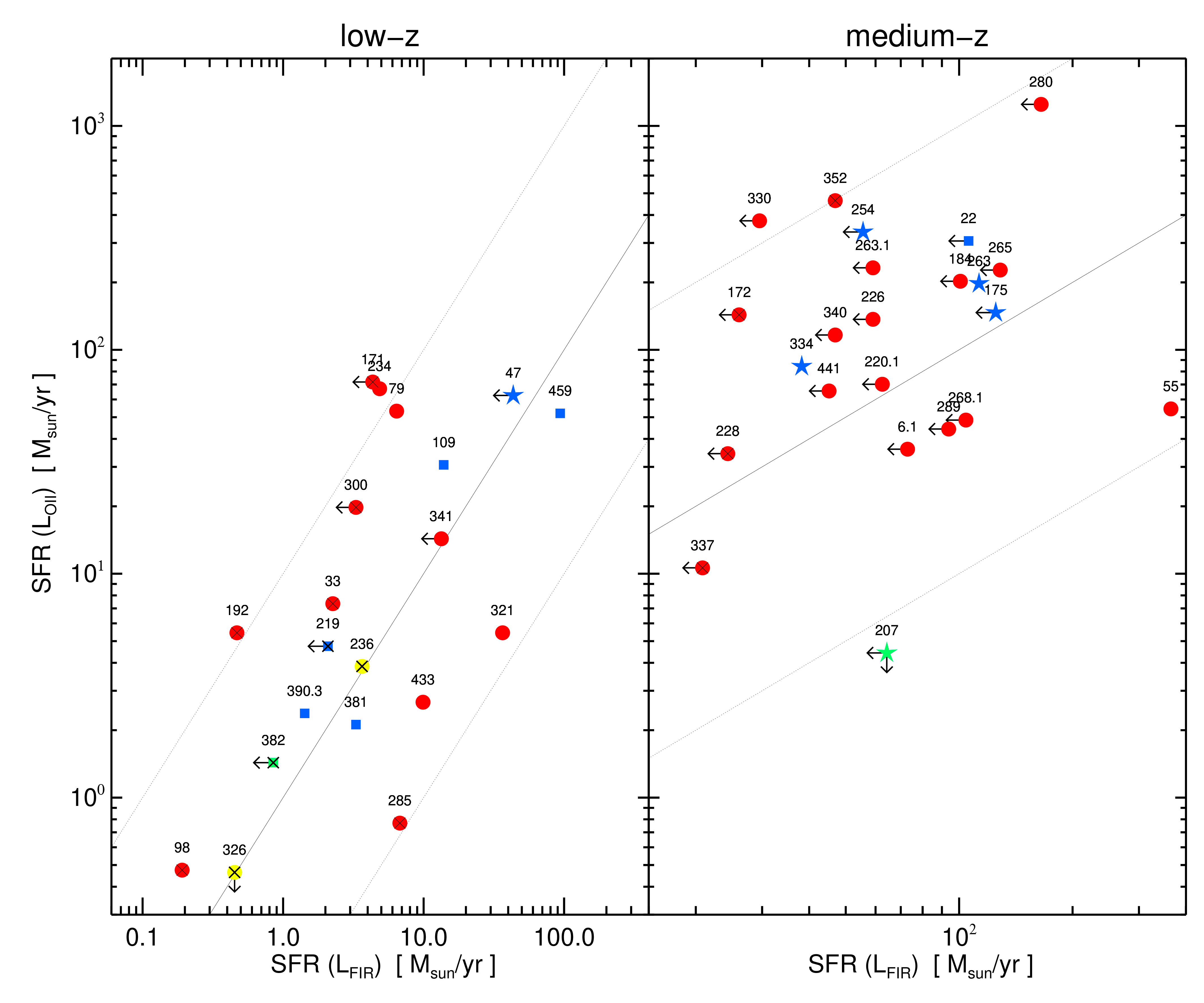}
  \caption{Star formation rate derived from [\ion{O}{2}] versus that
    from FIR for the low and medium-redshift sample (left/right
    panel).  $L_{\rm{O}\,II}$ is taken from \cite{Jackson97} and
    $L_{\rm {FIR}}$ from our modified blackbody fit.  The lines are
    not a fit; they illustrate the overall agreement with slope unity
    (solid) and a factor 10 above/below unity (dotted).  Notation as
    in Fig.~\ref{fig_result_mir_weak_med}. }
  \label{fig_result_OII_SFR_med}
\end{figure*}

To estimate the star forming luminosity $L_{\rm {FIR}}$ we assume that
the FIR emission fitted by the $\sim$30\,K modified blackbody is
entirely powered by stars. However, the AGN may heat dust at lower
temperatures than accounted for by the torus models of H\"onig et
al. (2010); in cases of different chemical dust composition and grain
geometries the AGN create larger emission at longer wavelengths
\citep{Sieb15}. Therefore $L_{\rm {FIR}}$ and the SFR derived from our
SED fits may be overestimated and the actual SFR may be
smaller. Therefore we treat our estimates as maximum possible SFR.
Other star formation indicators, e.g., via optical Balmer or [\ion{O}{2}] lines, may suffer even more from AGN contamination than the FIR. In fact there is evidence \citep{Hes93} that isotropic [\ion{O}{2}] emission from the narrow-line region is playing in important role in quasars and radio galaxies.

To provide a tentative cross check, we have converted 
both $L_{\rm {FIR}}$ and $L_{\rm{O}\,II}$ into star formation rates ${\rm SFR}_{\rm {FIR}}$ and
${\rm SFR}_{\rm{O}\,II}$ using the scaling relations (Eq. \ref{equ_sfr} and
\ref{equ_sfr_oii},  \citealt{Kennicutt98}). 

\begin{equation}
\label{equ_sfr_oii}
{\rm SFR} \left[\frac{M_{\odot}}{{\rm yr}}\right] = (1.4 \pm 0.4) \times 10^{-41} L_{\rm{O}\,II} \left[\frac{{\rm erg}}{{\rm s}}\right]
\end{equation}

The SFRs derived from both indicators match within an order of magnitude (Fig.~\ref{fig_result_OII_SFR_med}). There is no correlation between ${\rm SFR}_{\rm{O}\,II}$ and ${\rm SFR}_{\rm {FIR}}$ in the medium-redshift sample, where a higher AGN-triggered [\ion{O}{2}] emission has to be taken into account. For the low-redshift sample we find correlation but ${\rm SFR}_{\rm{O}\,II}$ is larger than ${\rm SFR}_{\rm {FIR}}$. Because we have not subtracted AGN contributions from $L_{\rm{O}\,II}$ we suggest that ${\rm SFR}_{\rm{O}\,II}$ is overestimated as well.

\begin{figure*}[t!]
  \includegraphics[width=\linewidth]{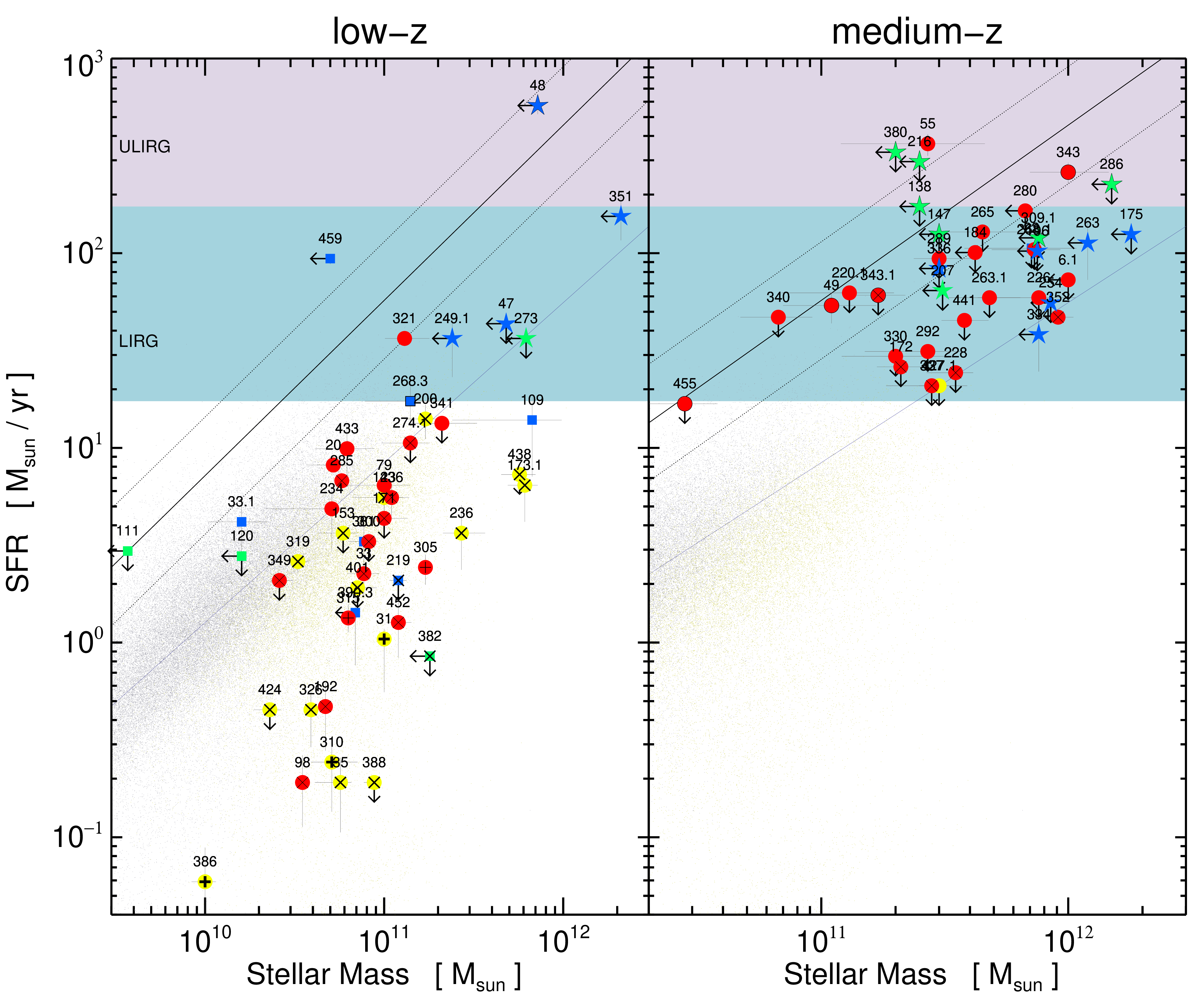}
  \caption{\label{fig_sf_double} Star formation rates, SFR vs
    stellar mass of the host for the low (left) and medium-redshift
    (right) samples.  The blue and violet shaded areas mark the range
    of LIRGs and ULIRGs.  The clouds of small grey and yellow dots
    indicate the main sequence of star forming galaxies and AGN,
    respectively, at low-redshift ($0.015 < z < 0.1$) from
    \cite{Brinchmann04} and \cite{Kauffmann03}) with the thin blue
    line representing the average relation of the star forming
    galaxies (\citealt{Elbaz07}). The thick black line flanked by the
    black dotted lines shows the average relation and 1$\sigma$ of the
    $0.8 < z < 1.2$ GOODS star forming galaxies from \cite{Elbaz07} .
    The location of the main sequence of star forming
    galaxies shifts up with increasing
    redshift \cite{Noeske07}. Notation as in Fig.~\ref{fig_result_mir_weak_med}.}
\end{figure*}

For most 3CR sources the ratio $L_{\rm {FIR}}/L_{\rm {Host}}<1$  is 
below that of the Milky Way.  This suggests that the
bulk of the sources contains a small amount of interstellar matter,
which serves as a gas reservoir for star formation.
To estimate the dust mass $M_{\rm D}$ from $L_{\nu}^{100~\micron}$ and the dust temperature
$T_{\rm D}$, we used 
%
%
%

\begin{equation}
  \label{equ_dust_mass}
  M_{\rm D}  = \frac{D_{{\rm L}}^2 F_{\nu}}{ \kappa_{\nu} B_{\nu} (T_{\rm D})}
\end{equation}
with dust opacity $\kappa_{100~\micron} = 27\,{\rm cm}^{2}/{\rm g}$
\citep{Draine03,Sieb14}.
The dust masses are listed in Tables \ref{3C_med_z_fir_fit} and \ref{3C_low_z_fir_fit}. 
For all but a handful starbursting 3CRs (identified below),  
the dust-to-stellar mass ratio lies in the range 
$10^{-3} - 10^{-5}$. 
That is 
about a factor 1--100 lower than for the Milky Way, i.e. more reminiscent of
dust-poor elliptical galaxies.
Despite large uncertainties in the dust mass estimates ($M_{{\rm D}}$ is especially sensitive to $T_{\rm{D}}$), this strongly
suggests that the bulk of the 3CRs contains only a
relatively small dust mass (and gas mass as well).
If the dust mass is widely distributed across the host galaxy, this 
results in a low dust column density, and
one may expect only a modest amount of 
overall optical extinction.
This may explain the rough agreement between ${\rm SFR}_{\rm{O}\,II}$ and ${\rm SFR}_{\rm {FIR}}$. 

In the following we use ${\rm SFR}_{\rm {FIR}}$. 
For comparison with other galaxy types and across
cosmic time,  
the derived SFR has to be placed in the context of the
already existing stellar mass $M_{\rm {Host}}$, derived from the
host luminosity via the intrinsic mass-to-light ratio of the synthetic
stellar population fits of the SEDs (Table
\ref{3C_med_z_host_fit}).
For FSQs $L_{\rm {FIR}}$ might be overestimated due to synchrotron
contamination and SFR upper limits are plotted. 
For type-2 AGN and steep-spectrum BLRGs, $M_{\rm {Host}}$ and SFR are not affected this way. 

For QSRs (SSQs and FSQs) the host luminosity is likely overestimated,
and we therefore plot upper limits for $M_{\rm {Host}}$ of these sources. 
We find some sources in the medium redshift sample, e.g. 3C\,343 or 3C\,6.1, with stellar masses up to $\approx 10^{12} M_{\odot}$, which are rare in the local universe ($z < 0.3$), but also have been reported recently for quasar host galaxies at redshifts $<$ 1 (see Fig.~7 of \cite{Matsuoka15}). For three sources (3C\,6.1, 3C\,184 and 3C\,280) there is no optical data available. In these sources the host fit is based on the WISE measurements where confusion within the WISE beam can not be excluded. Thus the host luminosity and derived stellar masses are treated as upper limits.



For the low-redshift sample the stellar masses lie in the range of
10$^{10}$ up to 10$^{12} M_{\odot}$, and the FIR luminosities are about
10$^{10} L_{\odot}$ (Tables~\ref{3C_low_z_host_fit} and
\ref{3C_low_z_fir_fit}). In Figure \ref{fig_sf_double} (left) the results are
plotted together with a comparison sample of SDSS galaxies in the
redshift range $0.015 < z < 0.1$ investigated by \cite{Brinchmann04}
and \cite{Kauffmann03}. The relation between SFR and mass for the SDSS galaxies is:

\begin{equation}
\label{equ_sfr_sdss}
{\rm SFR}_{{\rm SDSS}} \left[\frac{M_{\odot}}{{\rm yr}}\right] = 8.7 [-3.7, +7.4] \times \left[\frac{M_{\star}}{10^{11} M_{\odot}}\right]^{0.77}
\end{equation}

For the medium-redshift sample the range of $L_{\rm {FIR}}$ lies
around 10$^{11} L_{\odot}$ (Fig.\,\ref{fig_sf_double} right,
Table~\ref{3C_med_z_fir_fit}), i.e. ,that of IR luminous SF galaxies
(LIRGs) with a few sources above 10$^{12} L_{\odot}$ in the regime of
ULIRGs.  Derived host masses lie in the range of 10$^{11}$ -- 10$^{12}
M_{\odot}$, hence in the range of the most massive galaxies. For
comparison a selection of star forming galaxies from the GOODS fields
\citep{Elbaz07} with $10^9 M_{\odot} < M_{\star} < 10^{12}
M_{\odot}$, $1~M_{\odot}{\rm yr}^{-1} < {\rm SFR} < 300~M_{\odot}{\rm yr}^{-1}$), $0.8 < z < 1.2$ has:

\begin{equation}
\label{equ_sfr_goods}
{\rm SFR}_{\rm {GOODS}} \left[\frac{M_{\odot}}{{\rm yr}}\right] = 7.2 [-3.6, +7.2] \times \left[\frac{M_{\star}}{10^{10} M_{\odot}}\right]^{0.9}
\end{equation}


To provide a panoptic view, in Figure \ref{fig_sf_double} the whole
investigated 3CR sample is shown together with the appropriate
comparison samples.  The bulk of 3CR galaxies show only a small
specific star formation rate for their epoch.  The few exceptions are
3C\,49, 3C\,55 and 3C\,343 for the medium-redshift sample and 3C\,48,
3C\,321 and 3C\,459 for the low-redshift sample.


For the HERG, LERG and BLRG class, both stellar masses and star
formation rates  establish
that the 3Cs at $z < 1$ belong to the most massive galaxies of their epoch, but
most have low specific star forming activity. 
This is remarkably different from the results for 3CR sources at $1 < z < 2$ (\citealt{Podigachoski15a}), where $\approx$40\% show ULIRG-like SFRs (200-2000 $M_{\odot}$/yr) at similar host masses (10$^{11}$--10$^{12} M_{\odot}$).
This comparison suggests that many of the 3CRs at $z < 1$ are in a late evolutionary state. 
Alternatively, if their nuclear activity is triggered by galaxy
interactions/mergers, these could be dry mergers, i.e., the collision of two ISM-poor ellipticals.
Strikingly, the MIR-weak sources and LERGs  populate the lowest
end of the specific SFR distributions. 

\subsection{Evolution from HERGs over MIR-weak to LERGs?}
\label{sec_result_evolution}

The possibility that MIR-weak sources and LERGs may be considered as
classically accreting AGN in which the dust torus has a small covering
angle, and where a low extinction enables bright [\ion{O}{2}]
emission, and where the environment favours a high radio lobe
luminosity is raised in Sec. \ref{sec_result_mir_weak} and
\ref{subsec_result_mir_vs_radio_OIII}.
MIR-weak sources and LERGs populate the low end of the FIR
$R_{\rm {DR}}$ distributions (Fig.~\ref{fig_result_rdr030_med}).
The overall trends in Fig.~\ref{fig_result_rdr030_med}
make an evolutionary HERG-to-LERG
scenario attractiv, in which MIR-strong HERGs evolve to MIR-weak
HERGs and then further to LERGs. In this picture both the AGN
accretion luminosity as traced by the dust torus and the SF
luminosity (and the dust and gas mass) are high for young sources and decline
with increasing age.
However, there are several open issues with such a picture, as already
discussed by Ogle et al. (2006), who suggested that the LERGs are
jet-dominated sources with low, if any, accretion power.  Here we can
add more pieces to the puzzle:

{\rm { a)}} In the low-$z$ sample the LERGs and MIR-weak HERGs populate the same
host and
radio lobe luminosity ranges as the MIR strong HERGs and QSRs
(Figs.~\ref{fig_result_mir_weak_med} 
and~\ref{fig_result_mir_radio}). The same
holds for the lobe extent.
If accretion plays a substantial role in creating the radio power, one
would have to postulate that with declining accretion during the
HERG-to-LERG transition another power source grows, in order to keep the
lobe power similar for LERGs and HERGs. If accretion plays a subordinate role, then the jet power may be provided by other processes such as a spinning or a binary black hole in the nucleus.

{\rm { b)}} If the evolutionary HERG--LERG scenario is valid, one may expect
that it holds also at higher luminosities/redshifts. 
In the medium-$z$ sample, however, MIR-weak
sources (and LERGs) are rare, even with our new definition of 
MIR-weakness, and despite the denser circumgalactic medium compared to the local universe.
The five MIR-weak sources have the lowest lobe power among the medium-$z$ sample 
(Fig.~\ref{fig_result_mir_radio}).
This suggests that the MIR-weak successors of MIR-strong HERGs at
$0.5<z<1$ decline in lobe luminosity, so that they fall below the
178\,MHz flux limit of the 3CR sample. The time scales of radio
loudness are in the order of several 10$^{7}$ years, much shorter than
the age range of the sample.  Thus, any LERG-successors of MIR-strong
HERGs of the $0.5<z<1$ 3CR sample cannot have moved out of that
redshift range.  For example $t(z=0.55)-t(z=0.53)\approx1.5 \cdot
10^{8}\,{\rm yr}$ (\citealt{Wright06}).

{\rm { c)}} Another possibility is that MIR-weak
sources and LERGs do not exist in the early, say
$z>0.75$ universe and have recently evolved.
This may be related to the findings based on the Molonglo 2~Jy radio galaxies 
that LERGs are more frequently found in clusters than QSRs/BLRGs or HERGs (\citealt{Ramos13}).
Then the intra-cluster gas may enhance the radio lobes, 
as found for Cygnus A by Barthel \& Arnaud (1996).
Because the time scale for cluster formation is much higher than the radio-loud phase, 
a MIR-strong HERG in the field cannot evolve in several 10$^{7}$ years 
into a LERG surrounded by a cluster.

To summarize, there are several arguments against a simple evolutionary HERG-to-LERG
scenario, and much future work is required to solve the puzzle on the relation 
between the MIR-weak and MIR-strong sources.
 
\section{Summary and conclusions}
\label{chap_conclusions}

{\it {a) }} The $z <$ 0.5 sample contains many LERGs, the
medium-redshift sample many FSQs and in general more powerful
RGs/QSRs. Complete photometry from diverse catalogues was collected
for all sources to cover the SEDs continuously from optical to radio
wavelengths.

{\it {b) }} The new SEDs were used to measure host, AGN and star
forming luminosities by fitting appropriate templates with a
Metropolis-Hastings algorithm, based on maximizing the Bayesian
posterior probability. Also model-independent luminosities were
derived at selected wavelength, to quantify opacity effects and
non-thermal contributions from synchrotron emission.

{\it {c) }} The class of MIR-weak sources was investigated. A new
flux-ratio-dependent definition of MIR-weakness is given, which avoids
an absolute threshold. Compared to the previous definition, now
MIR-weak sources can clearly be separated also at higher
luminosities. Possible explanations of the MIR weakness are either an
extreme cool and thin dust torus or a lane seen directly from the
edge. The MIR-weak sources can also represent an intrinsically
different class of gas- and dust-poor AGN. Such a class may have
suffered from evolutionary depletion by an early strong merger history
in a more clustered environment.

{\it {d) }} The dust-to-radio-lobe luminosity ratios were calculated
in the range of $30-100~\micron$. This results in confirmation of the
unification hypothesis of HERGs and QSRs.

{\it {e) }} The sample reveals a decline of the radio-lobe--to--dust
luminosity ratio with increasing redshift.  This indicates a decline
of the efficiency to create radio lobes from early epochs up to today.
We suggest that this is caused by the dilution of the cirumgalactic
medium during cosmic evolution.

{\it {f) }} For the whole sample, stellar masses and star formation
rates were presented. This allows us to put the host galaxies of
radio-loud AGN into context with non-AGN and radio-quiet AGN at the same
epoch. The analysis shows that radio-loud AGN are associated with the
most-massive galaxies ($10^{10} \lesssim M_{\star} \lesssim 10^{12}$). In the majority of these galaxies new stars are
formed only at a low level. The SFR may even be smaller if the AGN-heated dust
torus contributes more at longer wavelength than indicated by
currently available models.

Altogether, the present Herschel observations of the 3CRs at $z<1$ do
not support the hypothesis that every radio-loud quasar is accompanied
by a high specific star forming activity.  Our analysis suggests that,
if radio-loud AGN are triggered by galaxy interactions
(Heckmann et al. 1986, Stockton et al. 1986), in most cases these are
probably dry mergers with little dust and gas mass.




\acknowledgments
\scriptsize

{\it Acknowledgments: }PACS has been developed by a consortium of institutes led by MPE (Germany) and including UVIE (Austria); KU Leuven, CSL, IMEC (Belgium); CEA, LAM (France); MPIA (Germany); INAF-IFSI/OAA/OAP/OAT, LENS, SISSA (Italy); IAC (Spain). This development has been supported by the funding agencies BMVIT (Austria), ESA-PRODEX (Belgium), CEA/CNES (France), DLR (Germany), ASI/INAF (Italy), and CICYT/MCYT (Spain).\\

SPIRE has been developed by a consortium of institutes led by Cardiff University (UK) and including Univ. Lethbridge (Canada); NAOC (China); CEA, LAM (France); IFSI, Univ. Padua (Italy); IAC (Spain); Stockholm Observatory (Sweden); Imperial College London, RAL, UCL-MSSL, UKATC, Univ. Sussex (UK); and Caltech, JPL, NHSC, Univ. Colorado (USA). This development has been supported by national funding agencies: CSA (Canada); NAOC (China); CEA, CNES, CNRS (France); ASI (Italy); MCINN (Spain); SNSB (Sweden); STFC, UKSA (UK); and NASA (USA).\\

This work is based in part on observations made with the Spitzer Space Telescope, which is operated by the Jet Propulsion Laboratory, California Institute of Technology under a contract with NASA.\\

This publication makes use of data products from the {\it Wide-field Infrared Survey Explorer}, which is a joint project of the University of California, Los Angeles, and the Jet Propulsion Laboratory/California Institute of Technology, funded by the National Aeronautics and Space Administration.\\

This publication makes use of data products from the {\it Two Micron All Sky Survey}, which is a joint project of the University of Massachusetts and the Infrared Processing and Analysis Center/California Institute of Technology, funded by the National Aeronautics and Space Administration and the National Science Foundation.\\

Funding for the creation and distribution of the SDSS Archive has been provided by the Alfred P. Sloan Foundation, the Participating Institutions, the National Aeronautics and Space Administration, the National Science Foundation, the U.S. Department of Energy, the Japanese Monbukagakusho, and the Max Planck Society. The SDSS Web site is http://www.sdss.org/.

The SDSS is managed by the Astrophysical Research Consortium (ARC) for the Participating Institutions. The Participating Institutions are The University of Chicago, Fermilab, the Institute for Advanced Study, the Japan Participation Group, The Johns Hopkins University, the Korean Scientist Group, Los Alamos National Laboratory, the Max-Planck-Institute for Astronomy (MPIA), the Max-Planck-Institute for Astrophysics (MPA), New Mexico State University, University of Pittsburgh, University of Portsmouth, Princeton University, the United States Naval Observatory, and the University of Washington. \\

This research has made use of NASA's Astrophysics Data System.\\

This research has made use of the VizieR catalogue access tool, CDS, Strasbourg, France. The original description of the VizieR service was published in \citealt{Ochsenbein00}.\\

This work is supported by the Nordrhein-Westf\"alische Akademie der
Wissenschaften und der K\"unste in the framework of the academy
program of the Federal Republic of Germany and the state
Nordrhein-Westfalen, and by Deutsches Zentrum f\"ur Luft-und Raumfahrt
(DLR 50\,OR\,1106). 

CW thanks BJW, JK and SPW for their hospitality during his 4 weeks visit at
the CfA, kindly granted by the Smithsonian Institution and Henrik Spoon for providing the IRS spectra.

We thank the referee for constructive comments.

\bibliographystyle{natbib}
\bibliography{biblio}

\vspace{-5mm}

\appendix

\section{Appendix material}

\subsection{Tables}
\label{app_tables}

\vspace{-5mm}

\begin{table}[h]
\scriptsize
\caption{Median SED composition at medium-redshift $0.5 < z < 1.0$\label{table_med_templ_seds}}
\begin{tabular}{ccccc}
\hline
&&&&\\
FSQs & SSQs & \multicolumn{3}{c}{HERGs}\\
&&&&\\
\hline
&&&&\\
3C138.0 &  3C022.0 &  3C049.0 &  3C263.1  & 3C172.0\\
3C147.0 &  3C175.0 &  3C055.0 &  3C289.0  & 3C220.1\\
3C207.0 &  3C196.0 &  3C184.0 &  3C330.0  & 3C228.0\\
3C216.0 &  3C254.0 &  3C226.0 &  3C455.0  & 3C268.1\\
3C286.0 &  3C263.0 &  3C265.0 &  3C340.0  & 3C292.0 \\
3C309.1 &  3C334.0 &  3C280.0 &  3C343.1  & 3C337.0 \\
3C380.0 &  3C336.0 &  3C343.0 &  3C441.0  & 3C352.0 \\
        &          &  3C006.1 &           & \\
&&&&\\         
\hline              
\end{tabular}       
\end{table}    

\vspace{-5mm}

\begin{table}[h]
\scriptsize
\caption{Median SED composition at low-redshift $z < 0.5$ (3C\,427.1 taken from medium-redshift sample)\label{table_low_templ_seds}}
\begin{tabular}{ccccccc}
\hline
&&&&&&\\
FSQs & SSQs & BLRGs & \multicolumn{2}{c}{HERGs} & \multicolumn{2}{c}{LERGs}\\
&&&&&&\\
\hline
&&&&&&\\
3C111.0 & 3C047.0 & 3C033.1 & 3C020.0 & 3C305.0 & 3C123.0 & 3C031.0\\
3C120.0 & 3C048.0 & 3C109.0 & 3C033.0 & 3C349.0 & 3C153.0 & 3C386.0\\
3C273.0 & 3C249.1 & 3C219.0 & 3C079.0 & 3C452.0 & 3C173.1 & 3C236.0\\
3C382.0 & 3C351.0 & 3C268.3 & 3C171.0 & 3C035.0 & 3C200.0 & 3C310.0\\
        &         & 3C381.0 & 3C234.0 & 3C192.0 & 3C319.0 & 3C388.0\\
        &         & 3C390.3 & 3C321.0 & 3C274.1 & 3C401.0 & 3C326.0\\
        &         & 3C459.0 & 3C341.0 & 3C300.0 & 3C424.0 & 3C427.1\\
        &         &         & 3C433.0 & 3C315.0 & 3C438.0 &        \\
        &         &         & 3C098.0 & 3C436.0 &         &        \\
        &         &         & 3C285.0 &         &         &        \\
&&&&&&\\         
\hline       
\end{tabular}  
\end{table}    

\vspace{-5mm}

\begin{table}[h!]
\caption{Model parameters\label{table_params}}
\scriptsize

\begin{tabular}{cccccccccccc}
\hline
\multicolumn{4}{l}{}          & \multicolumn{6}{l}{}           & \multicolumn{2}{l}{} \\
\multicolumn{4}{l}{Host} & \multicolumn{6}{l}{Torus} & \multicolumn{2}{l}{FIR} \\
\multicolumn{4}{l}{}          & \multicolumn{6}{l}{}           & \multicolumn{2}{l}{}\\
\hline
\hline

& & & & & & & & & & & \\
$A_{\rm {Host}}$ & Metal. & $\upsilon$ & Age & $A_{\rm {AGN}}$ & i & N & a & $\theta$ & $\tau$ & $A_{\rm {FIR}}$ & $T_{\rm {FIR}}$\\
&&&&&&&&&&&\\
 ({W}/{m$^2$}) &  & (log{yr})& (log{yr})& ({W}/{m$^2$}) & (deg)&  &  & (deg) & & ({W}/{m$^2$}) & (K)\\
&&&&&&&&&&&\\
\hline
&&&&&&&&&&&\\
var.\tablenotemark{a}& 0-1\tablenotemark{b} & 6.5-11 & 9-10.2 & var.\tablenotemark{a} & 0-90\tablenotemark{c} & 2.5-10\tablenotemark{d} & -2-0\tablenotemark{e} & 5-60\tablenotemark{f} & 30-80\tablenotemark{g} & var.\tablenotemark{a}& var.\tablenotemark{a}\\
&&&&&&&&&&&\\
\hline
\end{tabular}
\tablenotetext{a}{varies in the range of 200\% to one 50\% of individual start value}
\tablenotetext{b}{the interval of [0, 1] corresponds to metallicities of Z=[0.004, 0.008, 0.02, 0.05]}
\tablenotetext{c}{steps of 15 degree}
\tablenotetext{d}{steps of 2.5}
\tablenotetext{e}{steps of 0.5}
\tablenotetext{f}{steps of [5,30,45,60]}
\tablenotetext{g}{steps [30,50,80]}
\end{table}

\begin{table}  
 \scriptsize
   \caption{Host fit parameters for the 3Cs at $0.5 < z < 1.0$}
   \label{3C_med_z_host_fit}
   \centering
  \begin{tabular}{cccccccc}
  \hline
  &&&&&&&\\    
  Name & $A_{\rm {Host}} [log(\frac{W}{m^2})]$ & Z & $\upsilon$ [log(yr)] & Age [log(yr)] & $M_{\star} [10^{11} M_{\odot}]$ &  $L_{\rm {Host}} [10^{11} L_{\odot}]$ & $M_{\rm {BH}} [10^7 M_{\odot}]$\\
  &&&&&&&\\    
  \hline
  &&&&&&&\\    
   3C006.1 &          $-15.18^{+0.10}_{-0.11}$ &         0.008 &  $ 9.0^{+0.9}_{-0.9}$ &    $ 9.90^{+0.13}_{-0.16}$ &    $10.0^{+3.8}_{-3.9}$ &    $8.2^{+3.1}_{-2.1}$ &    $208.9^{+232.7}_{-123.3}$ \\
   3C022.0 &          $-15.18^{+0.10}_{-0.19}$ &          0.02 &  $10.0^{+0.3}_{-0.6}$ &    $ 9.50^{+0.09}_{-0.12}$ &     $7.1^{+3.7}_{-3.3}$ &   $11.0^{+3.1}_{-4.1}$ &     $142.4^{+188.3}_{-90.5}$ \\
   3C049.0 &          $-15.40^{+0.14}_{-0.14}$ &         0.008 &  $ 7.6^{+0.8}_{-0.7}$ &    $ 9.30^{+0.12}_{-0.06}$ &     $1.1^{+0.6}_{-0.4}$ &    $2.5^{+0.9}_{-0.7}$ &        $17.6^{+18.9}_{-9.0}$ \\
   3C055.0 &          $-15.21^{+0.18}_{-0.19}$ &         0.004 &  $ 8.1^{+1.6}_{-1.2}$ &    $ 9.50^{+0.07}_{-0.07}$ &     $2.7^{+1.9}_{-1.5}$ &    $6.0^{+3.1}_{-2.3}$ &       $48.2^{+72.6}_{-32.9}$ \\
\hline
  \end{tabular}
\end{table}

\begin{table}  
 \scriptsize
 \caption{Host fit parameters for the 3Cs at $z < 0.5$}
  \label{3C_low_z_host_fit}
  \centering
  \begin{tabular}{cccccccc}
  \hline
  &&&&&&&\\    
  Name & $A_{Host} [log(\frac{W}{m^2})]$ & Z & $\upsilon$ [log(yr)] & Age [log(yr)] & $M_{\star} [10^{10} M_{\odot}]$ &  $L_{Host} [10^{10} L_{\odot}]$  & $M_{\rm {BH}} [10^7 M_{\odot}]$\\
  &&&&&&&\\    
  \hline
  &&&&&&&\\    
   3C020.0 &          $-14.89^{+0.12}_{-0.15}$ &         0.008 &  $ 7.1^{+0.4}_{-0.4}$ &    $ 9.90^{+0.06}_{-0.08}$ &      $5.2^{+1.6}_{-1.5}$ &        $3.6^{+1.2}_{-1.1}$ &          $7.6^{+5.0}_{-3.2}$ \\
   3C031.0 &          $-12.21^{+0.04}_{-0.03}$ &          0.02 &  $ 6.6^{+0.2}_{-0.1}$ &    $ 9.50^{+0.01}_{-0.02}$ &     $10.0^{+0.3}_{-0.9}$ &       $13.0^{+1.1}_{-0.8}$ &         $15.8^{+4.8}_{-4.5}$ \\
   3C033.0 &          $-13.57^{+0.01}_{-0.02}$ &          0.02 &  $ 8.0^{+0.3}_{-0.4}$ &    $ 9.60^{+0.12}_{-0.11}$ &      $7.7^{+1.7}_{-1.5}$ &        $7.8^{+0.2}_{-0.3}$ &         $11.8^{+6.7}_{-4.2}$ \\
   3C033.1 &          $-14.82^{+0.11}_{-0.16}$ &         0.004 &  $ 7.8^{+0.3}_{-0.5}$ &    $ 9.10^{+0.08}_{-0.08}$ &      $1.6^{+0.6}_{-0.6}$ &        $5.3^{+1.7}_{-1.6}$ &          $2.0^{+1.4}_{-0.9}$ \\
\hline
  \end{tabular}
\end{table}

\begin{table}  
 \scriptsize
 \caption{AGN fit parameters for the 3Cs at $0.5 < z < 1.0$}
  \label{3C_med_z_agn_fit}
  \centering
  \begin{tabular}{ccccccccc}
  \hline
  &&&&&&&&\\    
      Name &  $A_{AGN} [log(\frac{W}{m^2})]$ & i [deg] & N & a & $\theta$ [deg] & $\tau$ & $L_{AGN} [10^{11} L_{\odot}]$ & $\frac{L_{AGN}}{L_{Edd}}$\\
  &&&&&&&&\\    
  \hline
  &&&&&&&&\\    
   3C006.1 &          $-15.08^{+0.05}_{-0.05}$ &     45 &    10.0 &    0.00 &     30 &   50 &    $9.2^{+1.3}_{-0.9}$ &      $0.0138^{+0.0246}_{-0.0079}$\\
   3C022.0 &          $-14.57^{+0.03}_{-0.07}$ &      0 &     2.5 &   -2.00 &     30 &   30 &   $49.0^{+4.6}_{-7.7}$ &      $0.1076^{+0.2156}_{-0.0685}$\\
   3C049.0 &          $-14.92^{+0.04}_{-0.07}$ &     15 &     5.0 &   -1.50 &     05 &   30 &    $7.0^{+0.6}_{-1.1}$ &      $0.1240^{+0.1506}_{-0.0736}$\\
   3C055.0 &          $-14.43^{+0.07}_{-0.10}$ &     15 &     5.0 &   -0.50 &     30 &   30 &   $30.0^{+5.6}_{-5.2}$ &      $0.1945^{+0.5339}_{-0.1303}$\\
\hline
  \end{tabular}
\end{table}

\begin{table}  
 \scriptsize
 \caption{AGN fit parameters for the 3Cs at $z < 0.5$}
  \label{3C_low_z_agn_fit}
  \centering
  \begin{tabular}{ccccccccc}
  \hline
  &&&&&&&&\\    
  Name &  $A_{AGN} [log(\frac{W}{m^2})]$ & i [deg] & N & a & $\theta$ [deg] & $\tau$ & $L_{AGN} [10^{9} L_{\odot}]$  & $\frac{L_{AGN}}{L_{Edd}}$\\
  &&&&&&&&\\    
  \hline
  &&&&&&&&\\    
   3C020.0 &          $-14.74^{+0.03}_{-0.07}$ &    45 &     5.0 &   -1.50 &    30 &   50 &        $52.0^{+7.6}_{-8.6}$ &      $0.0213^{+0.0211}_{-0.0106}$\\
   3C031.0 &          $-14.39^{+0.23}_{-0.27}$ &    45 &     5.0 &   -0.50 &    30 &   30 &         $0.7^{+0.5}_{-0.3}$ &      $0.0001^{+0.0002}_{-0.0001}$\\
   3C033.0 &          $-13.82^{+0.03}_{-0.04}$ &    60 &     7.5 &   -1.00 &    45 &   50 &        $42.0^{+3.6}_{-3.3}$ &      $0.0111^{+0.0077}_{-0.0046}$\\
   3C033.1 &          $-14.19^{+0.03}_{-0.04}$ &     0 &     7.5 &   -0.50 &    30 &   30 &     $210.0^{+15.0}_{-18.0}$ &      $0.3224^{+0.2865}_{-0.1458}$\\
\hline
  \end{tabular}
\end{table}

\newcommand{\promille}{%
    \kern-.05em%
    \raise.5ex\hbox{\the\scriptfont0 0}%
    \kern-.15em/\kern-.15em%
    \lower.25ex\hbox{\the\scriptfont0 00}}
\begin{table}  
 \scriptsize
 \caption{FIR fit parameters for the 3Cs at $0.5 < z < 1.0$\label{3C_med_z_fir_fit}}
  \centering
  \begin{tabular}{ccccccc}
  \hline
  &&&&&&\\    
  Name &  $A_{FIR} [log(\frac{W}{m^2})]$ &         $T_{FIR}$ [K] &    $SFR [\frac{M_{\odot}}{yr}]$ &   $L_{FIR} [10^{11} L_{\odot}]$ & $M_{Dust} [10^7 M_{\odot}]$ & $M_{Dust}$/$M_{\star} [\promille]$ \\
  &&&&&&\\    
  \hline
  &&&&&&\\    
   3C006.1 &          $-15.33^{+0.07}_{-0.08}$ &  $34.0^{+2.0}_{-2.0}$ &          $73.0^{+12.9}_{-12.5}$ &             $4.2^{+0.7}_{-0.7}$ & $11.58^{+0.17}_{-4.76}$ & $0.116$ \\	      
   3C022.0 &          $-15.28^{+0.13}_{-0.21}$ &  $31.0^{+0.4}_{-0.4}$ &         $106.0^{+36.5}_{-40.0}$ &             $6.1^{+2.1}_{-2.3}$ & $27.18^{+3.92}_{-7.15}$ & $0.383$ \\	      
   3C049.0 &          $-15.14^{+0.07}_{-0.09}$ &  $37.0^{+0.9}_{-0.9}$ &           $53.8^{+9.0}_{-10.2}$ &             $3.1^{+0.5}_{-0.6}$ & $4.07^{+0.42}_{-1.13}$ & $0.370$ \\	      
   3C055.0 &          $-14.48^{+0.05}_{-0.06}$ &  $37.0^{+0.5}_{-0.5}$ &         $364.8^{+43.4}_{-48.6}$ &            $21.0^{+2.5}_{-2.8}$ & $26.45^{+1.14}_{-3.82}$ & $0.980$ \\
\hline
  \end{tabular}
\end{table}

\begin{table}  
 \scriptsize
 \caption{FIR fit parameters for the 3Cs at $z < 0.5$\label{3C_low_z_fir_fit}}
  \centering
  \begin{tabular}{ccccccc}
  \hline
  &&&&&&\\    
  Name &  $A_{FIR} [log(\frac{W}{m^2})]$ &         $T_{FIR}$ [K] &    $SFR [\frac{M_{\odot}}{yr}]$ &    $L_{FIR} [10^{9} L_{\odot}]$ & $M_{Dust} [10^7 M_{\odot}]$ & $M_{Dust}$/$M_{\star} [\promille]$\\
  &&&&&&\\    
  \hline
  &&&&&&\\    
   3C020.0 &          $-14.66^{+0.05}_{-0.03}$ &  $22.0^{+0.3}_{-0.3}$ &             $8.2^{+1.1}_{-0.6}$ &            $47.0^{+6.1}_{-3.4}$ & $5.74^{+0.28}_{-0.63}$ & $1.103$ \\ 
   3C031.0 &          $-13.39^{+0.05}_{-0.27}$ &  $23.0^{+0.7}_{-0.7}$ &             $1.0^{+0.1}_{-0.5}$ &             $6.0^{+0.8}_{-2.8}$ & $0.90^{+0.07}_{-0.24}$ & $0.090$ \\ 
   3C033.0 &          $-14.19^{+0.07}_{-0.12}$ &  $30.0^{+0.6}_{-0.6}$ &             $2.3^{+0.5}_{-0.5}$ &            $13.0^{+2.6}_{-3.1}$ & $0.50^{+0.07}_{-0.16}$ & $0.065$ \\ 
   3C033.1 &          $-15.00^{+0.13}_{-0.21}$ &  $40.0^{+0.6}_{-0.6}$ &             $4.2^{+1.4}_{-1.5}$ &            $24.0^{+8.2}_{-8.9}$ & $0.33^{+0.02}_{-0.02}$ & $0.204$ \\
\hline
  \end{tabular}
\end{table}

\begin{table}  
 \scriptsize
 \caption{Monochromatic luminosities for the 3Cs at $0.5 < z < 1.0$}
  \label{3C_med_lum_mono}
  \centering
  \begin{tabular}{cccccc}
  \hline
  &&&&&\\
  Name &       $\nu L_{\nu}^{30GHz} [log(L_{\odot})]$ &   $\nu L_{\nu}^{178MHz} [log(L_{\odot})]$ &  $\nu L_{\nu}^{30~\micron} [log(L_{\odot})]$ &  $\nu L_{\nu}^{60~\micron} [log(L_{\odot})]$ & $\nu L_{\nu}^{100~\micron} [log(L_{\odot})]$ \\
  &&&&&\\
  \hline
  &&&&&\\
   3C006.1 &               $10.69 \pm 0.023$ &               $10.28 \pm 0.003$ &               $11.57 \pm 0.015$ &               $11.63 \pm 0.009$ &               $11.56 \pm 0.098$ \\
   3C022.0 &               $10.57 \pm 0.011$ &               $10.33 \pm 0.007$ &               $12.19 \pm 0.016$ &               $11.97 \pm 0.006$ &               $11.75 \pm 0.085$ \\
   3C049.0 &               $10.26 \pm 0.014$ &                $9.76 \pm 0.009$ &               $11.47 \pm 0.020$ &               $11.42 \pm 0.033$ &               $11.26 \pm 0.082$ \\
   3C055.0 &               $10.23 \pm 0.031$ &               $10.38 \pm 0.065$ &               $12.28 \pm 0.007$ &               $12.29 \pm 0.008$ &               $12.07 \pm 0.041$ \\
\hline
  \end{tabular}
\end{table}

\begin{table}  
 \scriptsize
 \caption{Monochromatic luminosities for the 3Cs at $z < 0.5$}
  \label{3C_low_lum_mono}
  \centering
  \begin{tabular}{cccccc}
  \hline
  &&&&&\\
  Name &       $\nu L_{\nu}^{30GHz} [log(L_{\odot})]$ &   $\nu L_{\nu}^{178MHz} [log(L_{\odot})]$ &  $\nu L_{\nu}^{30~\micron} [log(L_{\odot})]$ &  $\nu L_{\nu}^{60~\micron} [log(L_{\odot})]$ & $\nu L_{\nu}^{100~\micron} [log(L_{\odot})]$ \\
  &&&&&\\
  \hline
  &&&&&\\
   3C020.0 &                $9.65 \pm 0.010$ &                $9.16 \pm 0.003$ &               $10.46 \pm 0.049$ &               $10.40 \pm 0.066$ &               $10.24 \pm 0.014$ \\
   3C031.0 &                $7.21 \pm 0.058$ &                $6.58 \pm 0.009$ &                $8.98 \pm 0.157$ &                $9.31 \pm 0.001$ &                $9.57 \pm 0.045$ \\
   3C033.0 &                $8.73 \pm 0.034$ &                $8.24 \pm 0.002$ &               $10.38 \pm 0.028$ &               $10.17 \pm 0.046$ &                $9.94 \pm 0.096$ \\
   3C033.1 &                $9.05 \pm 0.041$ &                $8.72 \pm 0.002$ &               $11.01 \pm 0.033$ &               $10.68 \pm 0.003$ &               $10.29 \pm 0.001$ \\
\hline
  \end{tabular}
\end{table}

\clearpage
\subsection{Maps}
\label{app_maps}

No Maps available in preprint.

\end{document}